\documentclass[twocolumn,useAMS,usenatbib]{mnras}

\usepackage{color}
\usepackage{amsmath}
\usepackage{graphicx}
\usepackage{amssymb}
\usepackage{psfrag}

\usepackage{etoolbox}
\makeatletter
\patchcmd\@combinedblfloats{\box\@outputbox}{\unvbox\@outputbox}{}{%
   \errmessage{\noexpand\@combinedblfloats could not be patched}%
}%
 \makeatother

\newcommand{\beq}{\begin{equation}}
\newcommand{\eeq}{\end{equation}}
\newcommand{\bay}{\begin{array}}
\newcommand{\eay}{\end{array}}
\newcommand{\beqa}{\begin{align}}
\newcommand{\eeqa}{\end{align}}
\newcommand{\beqy}{\begin{eqnarray}}
\newcommand{\eeqy}{\end{eqnarray}}
\newcommand{\nn}{\nonumber}
\newcommand{\rmd}{\mathrm{d}}
\newcommand{\brac}[1]{\left({#1}\right)}
\newcommand{\pd}[2]{\frac{\partial{#1}}{\partial{#2}}}

\newcommand{\curl}{\nabla\times}
\renewcommand{\div}{\nabla\cdot}
\newcommand{\bj}{{\boldsymbol j}}
\newcommand{\bB}{{\boldsymbol B}}

\newcommand{\bE}{{\boldsymbol E}}

\newcommand{\bv}{{\boldsymbol v}}

\newcommand{\clM}{\mathcal{M}}

\newcommand{\maxw}{\boldsymbol{\mathcal{M}}}
\newcommand{\vpl}{\boldsymbol{v}_{\mathrm{pl}}}
\newcommand{\nhat}{\hat{\boldsymbol{n}}}

\newcommand{\skl}[1]{{\color{black}{#1}}}

\title[NS field evolution with plastic flow]{Magnetic-field evolution in a plastically-failing neutron-star crust}
\author[S. K. Lander \& K. N. Gourgouliatos]{S. K. Lander${}^1$\thanks{skl@camk.edu.pl} and
         K. N. Gourgouliatos${}^2$\thanks{konstantinos.gourgouliatos@durham.ac.uk}\\ \\
         ${}^1$Nicolaus Copernicus Astronomical Centre, Polish Academy of Sciences, Bartycka 18, 00-716 Warsaw, Poland,\\
         ${}^2$Department of Mathematical Sciences, Durham University, Durham, DH1 3LE, U.K.}

\begin{document}

\pagerange{\pageref{firstpage}--\pageref{lastpage}} \pubyear{0000}
\maketitle

\label{firstpage}

\begin{abstract}
Under normal conditions in a neutron-star crust, ions are locked in place in the crustal lattice and only electrons are mobile, and magnetic-field evolution is thus directly related to the electron velocity. The evolution, however, builds magnetic stresses that can become sufficiently large for the crust to exceed its elastic limit, and to flow plastically. We consider the nature of this plastic flow and the back-reaction on the crustal magnetic field evolution. We formulate a plane-parallel model for the local failure, showing that  surface motions are inevitable once the crust yields, in the absence of extra dissipative mechanisms. We perform numerical evolutions of the crustal magnetic field under the joint effect of Hall drift and Ohmic decay, tracking the build-up of magnetic stresses, and diagnosing crustal failure with the von Mises criterion. Beyond this point we solve for the coupled evolution of the plastic velocity and magnetic field. Our results suggest that to have a coexistence of a magnetar corona with small-scale magnetic features, the viscosity of the plastic flow must be roughly $10^{36}-10^{37}\ \textrm{g cm}^{-1}\textrm{s}^{-1}$. We find significant motion at the surface at a rate of $10-100$ centimetres per year, and that the localised magnetic field is weaker than in evolutions without plastic flow. We discuss astrophysical implications, and how our local simulations could be used to build a global model of field evolution in the neutron-star crust.
\end{abstract}


%%%%%%%%%%%%%%%%%%%%%%%%%%%%%%%%%%%%%%%%%%%%%%%%%%%%%%%%%%%%
%%%%%%%%%%%%%%%%%%%%%%%%%%%%%%%%%%%%%%%%%%%%%%%%%%%%%%%%%%%%
%%%%%%%%%%%%%%%%%%%%%%%%%%%%%%%%%%%%%%%%%%%%%%%%%%%%%%%%%%%%
\section{Introduction}

Many of the diverse phenomena observed from neutron stars (NSs) are believed to be related to magnetic-field evolution in their solid crusts. Most dramatically, the activity of magnetars -- bursts and flares in the hard X-ray/soft gamma-ray band, and their irregular enhanced spindown -- are most naturally explained in a paradigm where the solid crust has to \emph{fail}. The idea is that evolution of the star's magnetic field induces a secular build-up of stresses in the star's elastic crust, which eventually reaches its yield point; at this point there is a dynamical release of excess energy that heats the crust and produces coronal loops outside the star, in turn triggering emission processes that lead to the observed magnetar phenomena \citep{TD95,belo_thomp,belo09}.

Turning this picture into a quantitative model is, however, extremely challenging. It relies on detailed information about the star's internal magnetic-field strength and geometry, in both the core and the crust, and the processes by which the field evolves. It demands an understanding of how a neutron-star crust fails, both at a microscopic and macroscopic level, and the role played by the magnetic field in this failure. Finally, it requires knowledge of where the released elastic energy is transferred after crustal failure, and how this induces emission processes. No one could reasonably claim to have all of these ingredients under control yet, but there have been many recent promising developments.

Firstly, the emission part of the problem is arguably understood to a relatively satisfactory extent. Thermal blackbody emission is associated with the hot crust; persistent X-ray emission is likely related to coronal loops \citep{belo09}; timing changes and enhanced spindown to outer coronal regions \citep{belo_thomp}.

As for field evolution, if one restricts oneself to considering the magnetic field of a (never-yielding) crust in isolation from the core, the problem reduces to the simpler framework of electron MHD. This limiting case is so named because the ions are assumed fixed, and so the electric current (and consequently the magnetic field) depend only on one variable: the electron velocity through the crust. Magnetic-field evolution proceeds under the joint effect of the conservative Hall drift and the dissipative Ohmic decay. Electron MHD has been used extensively and successfully over the last decade or so to make quantitative studies of the magnetic-field evolution which takes place in a NS crust, and this field of study is now quite mature and well-explored. Building on the classic study of evolution mechanisms by \citet{GR92} and some early analytic work \citep{cumming04}, numerical studies -- in particular by Pons and collaborators -- have made steady progress towards full magneto-thermal evolutions in 2D \citep{ponsgepp07,aguilera,pons09,vigano13}. Recently, full 3D studies of the problem have appeared, demonstrating the radical differences between poloidal- and toroidal-dominated fields \citep{wood15,gour16,gour_holl}, along with 3D studies of the generalised neutron-star induction equation \cite{Vigano:2018}. \citet{akgun18} have made a first effort to connect the crustal evolution to the magnetosphere (and therefore to observable emission).

Core-field evolution is a much more vexed issue, mainly due to the uncertain role played by superfluid components (which are present even in magnetars younger than any observed to date; \citet{HGA}). There are debates about whether \emph{any} core process is fast enough to be relevant to magnetars; providing arbitration of the dispute is beyond the scope of this paper, but for a flavour of the recent literature see, e.g., \citet{jones06,GJS,graber,gug_alp_16,dommes,passa17,castillo,ofengeim}. There is also little understanding about how core and crust are linked, and the physics at the boundary that separates them. Even if core-field evolution is negligible, however, the magnetic field residing there is likely still important in providing the inner boundary condition for crustal-field evolution. In this case, one needs to know whether the majority of the core consists of type-I or type-II superconducting protons. In the latter case the microscopic field is quantised into thin fluxtubes \citep{BPP}, which may be averaged over to produce a macroscopic magnetic force \citep{EP77,mendell,GAS}, and the equilibrium equations may then be solved to find candidate macroscopic magnetic field configurations \citep{akgun_wass,L13,hen_wass,L14}. In the former case, the local field does not have a definite geometry, and it is less clear what form the macroscopic field would take \citep{sedrak05}.

Regardless of where the field evolution takes place, however, understanding the response of the crust to high stress is pivotal: partially because the build-up of stresses is inevitable, and exceeding the yield limit seems highly likely \citep{TD95,pernapons,L15,L16, Li:2016, Bransgrove:2018}, but also because models to explain magnetar emission rely on crustal motions \citep{belo09}. The wide range of magnetar activity with a characteristic $0.1$-second timescale (e.g. \citet{gogus01}), comparable with that for an elastic shear wave in the crust, also buttresses the argument that these phenomena are driven by a mechanism involving the transfer of crustal energy to the magnetosphere. The aim of this paper is to make the first quantitative study of the crust's coupled magneto-plastic dynamics, studying the plastic flow induced by an excess of magnetically-induced crustal stress, and the back-reaction of this flow on the magnetic field evolution.

%%%%%%%%%%%%%%%%%%%%%%%%%%%%%%%%%%%%%%%%%%%%%%%%%%%%%%%%%%%%
%%%%%%%%%%%%%%%%%%%%%%%%%%%%%%%%%%%%%%%%%%%%%%%%%%%%%%%%%%%%
%%%%%%%%%%%%%%%%%%%%%%%%%%%%%%%%%%%%%%%%%%%%%%%%%%%%%%%%%%%%
\section{Neutron-star field evolution with plastic flow}

\subsection{Magnetic-induced plastic flow}

\skl{To understand how magnetic stresses build in a neutron-star crust, and the crust's non-elastic response beyond its yield limit, we take an approach based on generalising the crust's momentum equation. This equation, whose form is effectively redundant in electron MHD, plays a key role in understanding the crust's behaviour at high stress. Our approach, outlined in this subsection, is a revised presentation of the approach in \citet{L16}.}

In its early life, a neutron star is entirely fluid, with a magnetic field developed during a variety of complex processes at birth. This field should complete its rearrangement into a stable hydromagnetic equilibrium state over a few Alfv\'en timescales -- a matter of seconds or less. This field is likely to be a dominantly large-scale structure dominated by low multipoles like the dipole component.  The crust begins to freeze very gradually over the following years, and finishes the process after a far longer time: roughly 1000 yr. Because this timescale is so much longer than the Alfv\'en one, the crust must freeze in a relaxed equilibrium state with the magnetic field; initially the crust harbours no stresses. The force balance is therefore given by
\beq \label{relaxed}
0 = -\nabla P_0-\rho_0\nabla\Phi_0+\div\maxw_0,
\eeq
where
\beq\label{maxwell}
\mathcal{M}_{ij}=\frac{1}{4\pi}\brac{B_i B_j-\frac{1}{2}B^2\delta_{ij}}
\eeq
is the Maxwell stress tensor, $P,\rho$ and $\Phi$ are, respectively, the pressure, mass density and gravitational potential; and the subscript zero indicates the value of quantities at the point when the crust freezes, and therefore when it is relaxed.
The divergence of $\maxw$ gives the familiar Lorentz force:
\beq
\div\maxw=\frac{1}{4\pi}(\bB\cdot\nabla)\bB-\frac{1}{8\pi}\nabla(B^2),
\eeq
where the first of the right-hand-side terms is the magnetic tension, and the latter the magnetic pressure.
Because we are studying a problem involving crustal stresses, however, we will often find it clearer to use $\maxw$ instead of $\bB$.

Over time the crustal magnetic field changes due to the action of Hall drift and Ohmic decay, and therefore evolves away from its relaxed state; the force balance now becomes
\beq \label{eom_eqm}
0 = -\nabla P-\rho\nabla\Phi+\div\maxw+\div\btau,
\eeq
where $\btau$ is the elastic stress tensor. This is still a stationary equilibrium state, but only because the crustal lattice inhibits motion, at the cost of building up stresses.

Let $\btau_{\mathrm{el}}$ be the maximum stress which the crust is able to sustain before failure. This is a tensorial quantity and need not be unique; various different combinations of the individual tensor components could lead to failure. \skl{Once $\btau_{\mathrm{el}}$ is exceeded, we will assume the crust deforms through an irreversible plastic flow; we discuss the physics of crustal failure and a quantitative failure criterion in the following subsection.} We will assume that any crustal failure is a simple shearing process; in this case the equation of motion for the crust gains an extra viscous-flow term corresponding to the piece of the elastic stress beyond the maximum value which the crust can sustain, and closely resembles the standard Navier-Stokes equation \citep{L16}:
\beq \label{eom_pl}
\skl{\rho\pd{\vpl}{t}+\rho(\vpl\cdot\nabla)\vpl}= -\nabla P-\rho\nabla\Phi+\div(\maxw+\btau_{\textrm{el}})+\nu\nabla^2\vpl,
\eeq
Within this paper, therefore, the `plastic flow' is physically the same as a standard viscous flow. Note, however, that if the failure geometry is less simple than the case we consider, the dynamics may depend explicitly on the (tensorial) rate of strain, through an extra term in the above equation \citep{prager}.

The crust is stably stratified, so that any compressible motions are inhibited by buoyancy forces. We therefore demand that the flow satisfies $\div\vpl=0$ (note that this is less restrictive than assuming constant density). We anticipate plastic flow in a
neutron-star crust to be slow and steady; slow meaning that we may
neglect the advective term $(\vpl\cdot\nabla)\vpl$ for being quadratic in the
velocity, steady meaning that we may neglect the acceleration term,
$\dot\bv_{\rm pl}=0$. Under these assumptions the plastic flow reduces to a standard Stokes flow, with
just the viscous term being important.

We do not expect major changes in the hydrostatic equilibrium state over time, and so assume that
\beq
P\approx P_0\ ,\ \Phi\approx\Phi_0.
\eeq
\skl{Strictly speaking, these neglected changes are formally of the same order of magnitude as the plastic flow itself -- quadratic in differences in the magnetic field. In proceeding, therefore, we should be aware that some small re-adjustment of the hydrostatic equilibrium could relieve some stresses and thus ultimately reduce the amount of energy transferred into the plastic flow compared with that predicted by our model.}
Note that if we had allowed for rotation, the corresponding centrifugal force term \emph{would} be expected to vary considerably over time, since spindown at magnetar-field strengths can be very rapid.

With the above assumptions, we first see from equations \eqref{relaxed} and \eqref{eom_eqm} that the elastic stress (below the yield point) is sourced by Maxwell stresses alone:
\beq\label{maxw_diff}
\btau=\maxw_0-\maxw.
\eeq
Next we compare the equations of motion above and below the yield stress,
\eqref{eom_pl} and \eqref{eom_eqm}; subtracting one from the other leads us to the (unsurprising) result that changes in the Lorentz force greater than those sustainable by the crust induce a viscous flow:
\beq \label{stokes_orig}
\nu\nabla^2\vpl = - \div(\maxw-\maxw_{\textrm{el}}),
\eeq
\skl{where $\maxw_{\textrm{el}}$ is the value of the Maxwell tensor immediately before the yield stress is attained -- recalling that this maximum tensorial stress is not uniquely defined.}
So far we have made concrete the notion of how magnetic stresses can eventually result in plastic motion of the crust, but have not described the mechanism of crustal failure in any detail. We will discuss this next, and in doing so will slightly revise the above equation.

\subsection{Failure criterion}
\label{fail_crit}

Arguably the biggest challenge in modelling any seismic phenomenon in a neutron star is that there are major uncertainties about how exactly the crystalline lattice of the crust yields at large stress. The terms `breaking' or `cracking' from older literature are misleading; at the high pressures relevant to a NS crust the material cannot break in the canonical fashion for terrestrial materials, in which a void propagates through the medium \citep{jones03,lev_lyu}. Microscopic simulations of crustal failure have provided valuable insight \citep{horo_kad,hoff_heyl}: they show that the crystal yields in a collective failure, undergoing an irreversible plastic deformation, but only once extreme levels of stress have built up (at least an order of magnitude higher than predicted by earlier work). Despite this result, we note that these simulations -- performed by a gradual shearing applied at the boundaries of the local domain -- may not unveil every failure mode of the system. It is also not clear whether crustal failure would proceed in the same way on larger scales. \skl{A failure event might appear collective on microscopic scales, whilst still being highly local with respect to the whole star. The plastically-flowing failed region may well be a wide patch or band -- but a narrow failure band between two large non-failing pieces of crust could mimic `cracking', notwithstanding the plastic nature of the localised failure.}

The notion of collective failure and the resultant plastic flow is difficult to capture in a large-scale model of the crust, since the boundaries of the region which fail may be irregular, set by impurities or the crust's seismic history. It is also not obvious what the `correct' boundary conditions would be for the failed region(s). In this paper we will sidestep this highly challenging issue by considering a local, plane-parallel, portion of the crust: macroscopic, but not large enough that we need to account for the curvature of the crust, nor for coexisting yielding and non-yielding regions within our numerical domain. We will start with a prescription that once \emph{any} gridpoint reaches failure criterion, the whole numerical domain -- i.e. the entire segment of the crust we model -- is assumed to fail. The plastic-flow velocity \eqref{stokes_orig} is then calculated everywhere (even though most of the numerical domain is below its yield stress), and the resultant additional term is evolved in the induction equation. This notion of collective failure is faithful to the physical picture of failure from previous work. However, it is clearly a rather extreme case, since in practice the crust is liable to fail first in the outermost, lowest-density region. 
As a second prescription, we will instead only begin to apply the additional, plastic-flow parts of the evolution once a gridpoint further into the crust reaches its local yield stress.  Fortunately, we will see later that our numerical results are rather insensitive to which presciption we use.

We will assume that failure occurs in the form of shearing motions, for a number of reasons. They satisfy our incompressibility condition; they result in the simplification of potentially complex viscoplastic dynamics to standard viscous dynamics (as discussed in the text around equation \eqref{eom_pl}); and they also allow us to adopt results from the aforementioned molecular-dynamics simulations with some degree of self-consistency (since they used shearing to induce failure). Furthermore, this class of motions are of most observational interest too, since they result in twisting of the magnetosphere and associated emission changes.

Terrestrial elasticity theory provides a variety of different relations used to predict yielding. A classic, simple yield criterion consistent with the physical picture we have for crust failing is that of von Mises, which is based on the concept that a material fails due to excessive shearing distortion rather than from volumetric changes. \citet{baiko_chug} argue that this criterion is not reliable for a perfect Coulomb crystal, but as long as our local domain is large enough that we may regard the material as polycrystalline, it should be safe to proceed. In order to employ the von Mises criterion, we must first separate the stress tensor into two pieces:
\beq
\tau_{ij}=\tilde\tau_{ij}+\eta\delta_{ij},
\eeq
where $\tilde\tau_{ij}$ and $\eta\delta_{ij}=\frac{1}{3}\tau_{kk}\delta_{ij}$ are, respectively, the deviatoric and volumetric pieces of the stress tensor. The former, trace-free, component is the relevant one in the von Mises criterion, which states that a medium yields when
\beq\label{vonmises_orig}
\tau_\mathrm{el}\leq\sqrt{\textstyle{\frac{1}{2}}\tilde\tau_{ij}\tilde\tau_{ij}}
= \sqrt{\textstyle{\frac{1}{2}}(\tilde\clM^0_{ij}-\tilde\clM_{ij}) (\tilde\clM^0_{ij}-\tilde\clM_{ij})}
\eeq
for some \emph{scalar} yield stress $\tau_{el}$ (this has no quantitative relation to the tensorial quantity $\btau_\mathrm{el}$ used in an indicative fashion in the earlier discussion).

In our model, elastic stresses build up due to changes in the Maxwell tensor \eqref{maxwell} alone, and so we need to find the deviatoric piece of this. We do so from its definition as the traceless piece of the full tensor:
\beq
\tilde{\mathcal{M}}_{ij}=\mathcal{M}_{ij}-\frac{1}{3}\mathcal{M}_{kk}\delta_{ij}=\frac{1}{4\pi}\brac{B_i B_j-\frac{1}{3}B^2\delta_{ij}},
\eeq
which differs from the full tensor only in the prefactor of the diagonal piece. Now, with the understanding that yielding of the crust is sourced by the deviatoric piece of the Maxwell stresses alone, we see that we must revise equation \eqref{stokes_orig} to read:
\beq
\nu\nabla^2\vpl = - \div(\tilde\maxw-\tilde{\maxw}_{\mathrm{el}}).
\eeq
This is now consistent with the failure criterion we adopt, but the numerical implementation of the above equation is problematic. In principle one can monitor elastic failure locally, by checking (at each timestep, and for each gridpoint) whether inequality \eqref{vonmises_orig} is satisfied. On the first timestep for which the inequality is satisfied at some gridpoint, one may record the components of the Maxwell stress deviator tensor to get the local value of $\tilde\maxw_{\textrm{el}}$. Over time more gridpoints will reach the yield stress, and the results recorded to build up an array for $\tilde\maxw_{\textrm{el}}$. However, in practical terms we cannot allow individual gridpoints to fail, since all the surrounding gridpoints provide a boundary condition to any motion and it would be physically improbable and numerically impossible to deal with; instead we expect larger regions to fail collectively. Nor can we complete building up the $\tilde\maxw_{\textrm{el}}$ array before we allow any failure to occur, because the outer region would have vastly exceeded the breaking stress before anything occurred near the inner boundary. Instead, for the sake of definiteness, we will use the relaxed-state $\tilde\maxw_0$ in lieu of $\tilde\maxw_{\mathrm{el}}$, keeping in mind that doing so will likely result in an overestimated plastic flow. Having done so, our final form of the equation sourcing the plastic flow is
\begin{align}
 4\pi\nu\nabla^2\vpl &= - 4\pi\div(\tilde\maxw-\tilde{\maxw}_0)\nn\\
  &=(\bB_0\cdot\nabla)\bB_0-(\bB\cdot\nabla)\bB-\frac{1}{3}\nabla(B_0^2-B^2).
 \label{stokes_new}
\end{align}
One might have guessed that the plastic flow would depend only on differences in magnetic tension; the above relation shows that the changing magnetic pressure is also important.
We may now also write the von Mises criterion in a more explicit form:
\beq\label{vonmises_mag}
\tau_\mathrm{el}\leq\frac{1}{4\pi}\sqrt{\textstyle{\frac{1}{3}}B_0^4+\textstyle{\frac{1}{3}}B^4+\textstyle{\frac{1}{3}}B_0^2 B^2-(\bB\cdot\bB_0)^2}.
\eeq
The prefactors of these terms differ from those in \citep{L15}, which calculated a failure criterion using the full Maxwell stress tensor (thus implicitly allowing crustal yielding to be sourced in part by the expansion/contraction of the crust). In the context of that paper the calculation was reasonable, since it was not concerned with what happens to the crust \emph{after} yielding. Here, however, we have had to be more explicit about the failure mechanism, so that our calculation of crustal failure is consistent with the resulting plastic flow.

\subsection{Electron MHD with plastic flow}

\skl{To understand the evolution of the crustal magnetic field in the presence of a plastic flow, we need to revise the usual derivation leading to electron MHD \citep{GR92}. We begin with Faraday's equation:
\beq
\pd{\bB}{t}=-c\curl\bE,
\eeq
where $\bE$ is the electric field and $c$ the speed of light. Now, electron MHD assumes} that the advection of the magnetic field lines is due to the motion of free electrons, which may be thought of as the only carriers of the electric currents in the crust (since the ions, locked into the crustal lattice, are static). Thus, the electron velocity can be related to the magnetic field via Amp\`ere's law:
\beq
\bv_{e}=-\frac{\bj }{e n_{e}}=-\frac{c}{4\pi e n_e} \nabla \times \bB \,,
\eeq
where $n_e$ is the electron number density and $e$ the elementary electron charge. 
We can then evaluate the electric field through Ohm's law:
\beq
\bE =-\frac{\bv_e \times \bB}{c} +\frac{\bj}{\sigma}\,,
\eeq
where $\sigma$ is the electric conductivity of the crust. Finally, we substitute into Faraday's induction equation to obtain the usual electron MHD equation:
\beq
\frac{\partial \bB}{\partial t}=-\nabla \times \left(\frac{c}{4 \pi e n_{e}} \left(\nabla \times \bB \right)\times \bB +\frac{c^2}{4 \pi \sigma} \nabla \times \bB \right)\, .
\eeq
The electron MHD equation holds provided the crust is strong enough to balance the force terms. Because of this, the momentum equation that appears in standard MHD (see equation \eqref{eom_eqm}), is trivially satisfied. Should part of crust fail and consequently initiate a plastic flow, we need to take into account its effect on the overall evolution. \skl{We will associate the plastic flow with the motion of the ions alone, and make the approximation that the electrons are not themselves advected by the flow; for this, we require $|\vpl|\ll |\bv_e|$. The new effect is that the electric current is no longer directly linked to $\bv_e$ alone, but takes its more usual form as being due to a relative flow of oppositely-charged fluids:
  \beq
\bj=-e n_e (\bv_e-\vpl).
  \eeq
Ohm's law -- with or without the slow plastic flow -- follows from the momentum equation of the electron momentum equation \citep{GR92}. What changes in the case of plastic flow, is that eliminating $\bv_e$ in favour of $\bj$ now picks up a new $\vpl$ term not present before. As a result Ohm's law becomes:
\beq
\bE =\frac{1}{c}\brac{\frac{\bj\times\bB}{e n_e} -\vpl\times\bB} +\frac{\bj}{\sigma}\,.
\eeq
}
Consequently the electron MHD equation with plastic flow becomes:
\begin{align}
\frac{\partial \bB}{\partial t}=-\nabla \times \bigg( &\frac{c}{4 \pi e n_{e}} \left(\nabla \times \bB \right)\times \bB -\vpl \times \bB \nonumber \\
&+\frac{c^2}{4 \pi \sigma} \nabla \times \bB \bigg)\, .
\label{EMHD_PLASTIC}
\end{align}

\subsection{Material properties of the crust}
\label{CRUST_PROPERTIES}
To solve the equations above, we need a model for the structure of a neutron-star crust and information about a number of its transport properties.
One key quantity is the electron number density $n_e$, which is given by
\beq\label{ne_eqn}
n_e=\frac{Z}{A}(1-x_{fn})\frac{\rho}{m_b},
\eeq
where $Z$ is the number of protons per ion, $A$ the number of baryons per ion, $x_{fn}$ the fraction of `free' neutrons outside ions and $m_b$ the baryon mass (more specifically, we use the value of the atomic mass unit here). $Z,A$ and $x_{fn}$ are all functions dependent on the density alone, so first we need $\rho$. We obtain this by solving the TOV equations for a hydrostatic equilibrium configuration \citep{MTW}, using the equation of state of \citet{DH01} based on the SLy4 interaction. We choose a configuration with a mass of $1.4$ solar masses, for which the stellar radius is $11.7$ km, the neutron-drip point is at $11.3$ km (i.e. the radius $R_{nd}$ of the density contour $\rho=\rho_{nd}\equiv 4\times 10^{11}\textrm{g cm}^{-3}$) and the crust-core boundary at $10.8$ km (i.e. the radius $R_{cc}$ of the density contour $\rho=\rho_{cc}\equiv 1.4\times 10^{14}\textrm{g cm}^{-3}$).

In the bulk of the literature on crustal dynamics, it is naively assumed that the relevant region of the star is the entire region for which $\rho<\rho_{cc}$. However, the outer crust ($\rho<\rho_{nd}$) only crystallises for temperatures below around $10^9$ K \citep{haen_pich,krueger} and so all or part of this region will be liquid for young neutron stars, or for those with an internal heating mechanism like magnetars \citep{kaminker}. Furthermore, even if parts of the outer crust are formally just cold enough to solidify, the crystalline lattice would presumably be so weak that it would not be able to harbour any significant stress. For these reasons, in this paper we consider magnetic-field evolution in \emph{the inner crust only}, with the outermost layer assumed to respond passively to this evolution, like a magnetosphere. It is therefore convenient for us to define a rescaled radius variable
\beq
\mathcal{R}\equiv \frac{(r-R_{cc})}{(R_{nd}-R_{cc})},
\eeq
which is zero at the crust-core boundary and unity at the neutron-drip point.

Our numerical solution for $\rho$ is itself accurate, but as we will see later, some transport properties we require for this work are far more poorly understood. Therefore, rather than adopting the exact output from our calculation of $\rho$, we will use the following simple approximation:
\beq
\tilde\rho\equiv\frac{\rho}{\rho_{cc}}=410\brac{1+\frac{R_{cc}}{R_{nd}}}^2(1-\mathcal{R})^2+0.004,
\eeq
which deviates little from the exact result in the region $\rho_{nd}<\rho\leq\rho_{cc}$.

\citet{DH01} provides tables of $Z,A,x_{fn}$ and $\rho$ throughout the crust. We use these results to construct polynomial fits to the former three quantities in terms of $\rho$, and from these quantities and equation \eqref{ne_eqn} we calculate the electron number density $n_e$. The calculated result may be approximated to good accuracy with the following simple expression, which we use in place of the full numerical result:
\beq
n_e=10^{36}(1.5\tilde\rho^{2/3}+1.9\tilde\rho^2)\textrm{ cm}^{-3}.
\eeq
Whilst the crust remains below its yield stress, the equations of electron MHD apply, for which we only need to know one transport coefficient: the electrical conductivity $\sigma$. In this work we will assume $\sigma \propto n_{e}^{2/3}$, and set the conductivity at the base of the crust to $\sigma(R_{cc})=10^{24}$s$^{-1}$. 
Next, the Coulomb parameter is given by
\beq
\Gamma=\frac{Z^2 e^2}{a_I k_B T}
\eeq
where $a_I=(4\pi n_I/3)^{-1/3}$ is the ion sphere radius. This quantity is of fundamental importance: it tells us when it is energetically favoured for the outer envelope of the neutron star to crystallise. In particular, a lattice forms when $\Gamma>175$.

For the breaking stress we use the fit of \citet{CH10} to their molecular-dynamics simulations of NS crust-yielding:
\beq
\tau_{el}=\brac{0.0195-\frac{1.27}{\Gamma-71}}\frac{Z^2 e^2 n_I}{a_I}.
\eeq
It is again convenient to use a fit to these results instead of the full output. We adopt the following approximation:
\beq
\tau_{el}=5.1\times 10^{29}(0.4\tilde\rho+0.5\tilde\rho^3)\textrm{ g\ cm}^{-1}\textrm{s}^{-2}.
\eeq
Finally, we need an expression for the viscosity $\nu$ of the crust in its plastic phase. This is essentially completely unknown from first principles, and so we will begin by using the estimate from \citet{L16}, which comes from demanding that a magnetar's corona can in principle be a persistent rather than a transient phenomenon. In other words, the viscosity must be low enough that crustal motions can twist the magnetosphere on a shorter timescale than the $10$-yr timescale on which exterior electric current is thought to dissipate \citep{belo_thomp}. This reasoning leads to a plastic viscosity of the same functional form as that for $\tau_{el}$, with only a different magnitude:
\beq
\nu=1.6\times 10^{38}(0.4\tilde\rho+0.5\tilde\rho^3)\textrm{ g\ cm}^{-1}\textrm{s}^{-1}.
\label{PLASTIC_NU}
\eeq
Note that a calculation of this quantity from theoretical grounds has been attempted by \citet{kwang-hua}, and is orders of magnitude lower than the above. If this were correct, plastic evolution would be so fast that the crust would be immediately returned to its elastic regime, and it seems doubtful that any large-scale gradual magnetospheric twisting could occur.

\begin{figure}
\begin{center}
\begin{minipage}[c]{\linewidth}
\includegraphics[width=\textwidth]{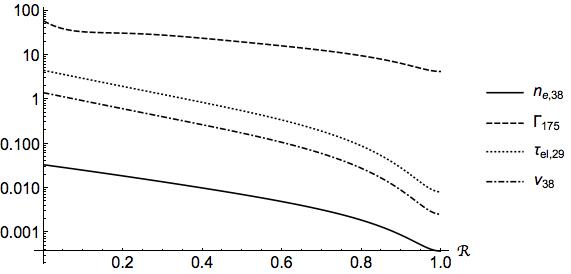}
\end{minipage}
\caption{\label{crust_qty}
  Key properties of our crustal model as a function of dimensionless radius $\mathcal{R}$, so that $\mathcal{R}=0$ and $1$ correspond to the crust-core and neutron-drip boundaries, respectively. Plotted are the electron number density $n_{e,38}$ in units of $10^{38}\textrm{cm}^{-3}$; the Coulomb parameter $\Gamma_{175}$ divided by 175 (the value at which crystallisation occurs); the breaking stress $\tau_{el,29}$ in units of $10^{29}\textrm{g\ cm}^{-1}\textrm{s}^{-2}$ and the viscosity of the crust's plastic phase $\nu_{38}$ in units of $10^{38}\textrm{g\ cm}^{-1}\textrm{s}^{-1}$.}
\end{center}
\end{figure}

\subsection{Boundary conditions}

We have argued above that the natural outer boundary for our work is the boundary between the inner and outer parts of the crust, defined by the neutron-drip isopycnic contour where $\rho\approx 4\times 10^{11}$ g cm${}^{-3}$. We will assume that this boundary is free for both the magnetic field and the plastic velocity, and implement this by demanding that the radial derivatives of both quantities vanish there (rather than the quantities themselves).  The magnetic field beyond this point is matched to a vacuum field, and any dynamics of the outer crust are assumed to be slave to the evolution occurring at greater depths.

The inner boundary is the crust-core boundary, at which we assume the radial magnetic magnetic field and plastic velocity to vanish. For the velocity this is quite natural; the stronger lattice at high densities is less susceptible to fail anyway, and if it does the motion will be much slower, due to the higher plastic viscosity there. The magnetic-field condition is standard and given little attention, but is harder to justify; the pragmatic reason to adopt it is to ensure that is no coupling between the crust and the (poorly-understood) magnetic-field evolution at the crust-core boundary and into the core. Since timescales for core evolution tend to be considerably longer than those of the crust, however, there is some reason to believe that the crustal evolution would be somewhat decoupled from core processes. Setting the radial field component to zero is one way to ensure this, albeit by brute force. For the tangential field component, one could demand that the tangential electric field vanishes, in accordance with the `Meissner boundary condition' described in \citet{Hollerbach:2002}. However, this leads to the inversion of a non-linear equation, drastically complicating the calculation and only leading to minimal difference from the straightforward zero-boundary condition that we implement here.

\subsection{Energy conservation and surface motions}

Three different effects drive the magnetic-field evolution we consider here, and may lead to changes in the star's total magnetic energy
\beq
E_{\textrm{mag}}=\frac{1}{8\pi}\int B^2\ \rmd V.
\eeq
The properties of two of these effects are well known: Ohmic decay results in the dissipation of magnetic energy, whilst the Hall drift effect is conservative \citep{GR92}. The behaviour of the final term, due to plastic flow, is less clear. We know, from equation \eqref{stokes_new}, that the plastic flow is sourced from the unbalanced Lorentz force in a process involving the plastic-phase viscosity $\nu$ -- but despite its connotations of dissipation, here $\nu$ acts more to regulate the transfer of energy from magnetic to (plastic) kinetic, rather than to dissipate it.

To gain some intuition for the effect of plastic flow on the total crustal energy, let us first consider a different but related effect: that of the dynamical term in the usual induction equation for MHD (i.e. in a fluid, not an elastic solid). With this term alone, the induction equation reads
\beq\label{vcrossB}
\pd{\bB}{t}=\curl(\bv\times\bB),
\eeq
and under the action of this term alone the magnetic energy of the system evolves as follows:
\beq
\dot{E}_{\textrm{mag}}=\frac{1}{4\pi}\int\bB\cdot\dot\bB\ \rmd V=\frac{1}{4\pi}\int\bB\cdot[\curl(\bv\times\bB)]\ \rmd V
\eeq
where we have used the product rule. This term does not appear to conserve magnetic energy, but this is not surprising: it describes a dynamical coupling between $\bv$ and $\bB$ which leads to (e.g.) Alfv\'en waves. Let us, then, instead check whether the term preserves \emph{total} energy. In the simplest possible set-up, assuming the velocity evolves in a way independent of the stellar pressure and gravity, we have
\beq\label{simple_euler}
\pd{\bv}{t}=\frac{1}{4\pi\rho}(\curl\bB)\times\bB
\eeq
and changes in the total energy should be given by the sum of changes in the kinetic and magnetic pieces:
\begin{align}
\dot{E}_{\textrm{tot}}&=\int\brac{\frac{1}{4\pi}\bB\cdot\dot\bB+\rho\bv\cdot\dot\bv}\ \rmd V\nn\\
                      &=\frac{1}{4\pi}\int\big\{\div[\bB\times(\bv\times\bB)]+\bv\cdot[\bB\times(\curl\bB)]\nn\\
                       &\hspace{2cm} +\bv\cdot[(\curl\bB)\times\bB]\big\}\ \rmd V,
\label{Edot_vdyn}
\end{align}
where the first two terms in the final equality come from the magnetic-energy piece, upon using a couple of vector identities, and the final piece comes from substituting in equation \eqref{simple_euler} for $\dot\bv$. Clearly the second and third terms cancel, leaving us (after using the divergence theorem) with a surface integral:
\beq
\dot{E}_{\textrm{tot}}=\int[(\bv\times\bB)\times\bB]\cdot\nhat\ \rmd S
\eeq
where $\nhat$ is the unit outward-pointing normal vector to the surface. So, if there are no losses from the surface due to electromagnetic waves, then the dynamical term of equation \eqref{vcrossB} is conservative.

Returning to the plastic-flow problem, as we have formulated it, we see that there is no such cancellation of terms. We are assuming that the flow is steady on the timescales of interest,  $\dot\bv_{\rm pl}=0$, meaning that the kinetic energy is roughly constant and only the magnetic energy is time-varying:
\beq
\dot{E}_{\textrm{mag}}=\frac{1}{4\pi}\int\bB\cdot[\curl(\vpl\times\bB)]\ \rmd V.
\eeq
Comparing with equation \eqref{Edot_vdyn}, we see that this may be written as
\beq
4\pi\dot{E}_{\textrm{mag}}
 = \int\big\{\div[\bB\times(\vpl\times\bB)]+\vpl\cdot[\bB\times(\curl\bB)]\big\}\ \rmd V,
\eeq
where the second term gives the \emph{power} of the plastic-flow process: the work done by the Lorentz force at a rate set by the plastic-flow velocity. From the above, we see that plastic flow is neither conservative or dissipative, a priori. For it to be conservative, we require (using another vector identity) that:
\beq
\int[(\vpl\times\bB)\cdot(\bB\times\nhat)\ \rmd S=-\int\vpl\cdot[\bB\times(\curl\bB)]\ \rmd V.
\eeq
We see that the power of the process -- involving the interior plastic flow calculated from equation \eqref{stokes_new} -- must source a surface plastic flow that satisfies the above relation.

Clearly it is not possible to satisfy the above relation unless $(\vpl\times\bB)$ has some component parallel to $(\bB\times\nhat)$ at the surface. By the definition of $\nhat$ we know that $(\bB\times\nhat)$ is tangent to the surface. For $(\vpl\times\bB)$ to have a tangential component, either $\vpl$ or $\bB$ must have a component parallel to $\nhat$. In other words, necessary -- but not sufficient -- conditions for conservative magnetic-field evolution under plastic flow are:
\beq
\vpl\big|_{\textrm{surface}}\neq 0\textrm{ and }\bB\big|_{\textrm{surface}}\neq 0
\eeq
\emph{and}:
\beq\label{vpl_conserv}
\textrm{either }\nhat\cdot\vpl\big|_{\textrm{surface}}\neq 0 \textrm{ or }\nhat\cdot\bB\big|_{\textrm{surface}}\neq 0.
\eeq
The most important implication of this is that once a neutron star crust yields under magnetic stresses, \emph{it must generate surface motions} in order to conserve energy. If there is no surface motion, all of the power must be expended in some other way -- presumably through viscoplastic heating in the crust.

%%%%%%%%%%%%%%%%%%%%%%%%%%%%%%%%%%%%%%%%%%%%%%%%%%%%%%%%%%%%
%%%%%%%%%%%%%%%%%%%%%%%%%%%%%%%%%%%%%%%%%%%%%%%%%%%%%%%%%%%%
%%%%%%%%%%%%%%%%%%%%%%%%%%%%%%%%%%%%%%%%%%%%%%%%%%%%%%%%%%%%
\section{Numerical implementation}

We numerically integrate equation Eq.~\eqref{EMHD_PLASTIC} (and, above the yield stress, equation \eqref{stokes_new}) in Cartesian geometry with orthogonal coordinates $(x,y,z)$. We consider a square block in $y-z$ of size $y\in [0, R_{nd}-R_{cc}]$ and $z\in [0, R_{nd}-R_{cc}]$, where $z$ is the depth of the crust and $z=0$ corresponds to the crust-core boundary. The system has plane-parallel geometry with the $x-$derivatives of the physical quantities vanishing. For this purpose we are using a modified version of a code previously developed and used for electron MHD evolution in \citet{Gourgouliatos:2015b, Gourgouliatos:2016b}.

Some classes of magnetic field do not evolve under the Hall effect at all; these are known as \emph{Hall equilibria} \citep{gour13}. In order to ensure our initial conditions cause immediate evolution, we take care to ensure that they are deliberately out of Hall equilibrium. This initial field corresponds to the crust-freezing state, so that although it has non-vanishing Maxwell stresses $\maxw_0$, these are in hydromagnetic equilibrium, and so the crust is unstressed with them. As the magnetic field evolves, we monitor the von Mises criterion (Eq.~\eqref{vonmises_mag}) to see whether it is satisfied at some location within the integration domain. Since we are considering a block of the crust, we assume that if the crust fails at some point this will initiate a global plastic flow within the block. To find the plastic flow velocity we then solve Eq.~\eqref{stokes_new}.%, and used in the numerical integration of Eq.~\ref{EMHD_PLASTIC}. 

%As the solution of the plastic flow velocity can slow substantially the simulation we perform this every $XXX$ number of steps. We have experimented with varying the frequency and we found that at the given resolution this provides accurate results. 

\subsection{Plane-parallel form of equations}

We express the magnetic field in terms of two scalars
\beq
\bB = B_x (y,z) \hat{\bf x}+  \nabla \Psi(y,z) \times \hat{\bf x}\,.
\eeq
This expression ensures that the magnetic field is divergence-free and incorporates its plane-parallel symmetry. The plastic flow is incompressible; $\nabla \cdot \bv_{pl}= 0$. Moreover, radial motions are energetically disfavoured, and thus we take the $z-$component of the velocity to be identically zero. Finally, the  plane-parallel assumption we have made leads to $\partial_x\bv_{pl}={\bf 0}$. Subject to these constraints the plastic velocity can be written in the following form
\beq
\bv_{pl}=v_{x}(y,z)\hat{\bf x}+ v_{y}(z) \hat{\bf y}\,.
\eeq
Next we rewrite the electron MHD equation with plastic evolution (Eq.~\ref{EMHD_PLASTIC}), which reduces to two scalar partial differential equations:
\beq
\partial_t \Psi=\frac{c}{4 \pi e n_{e}} \left(\nabla B_{x} \times \hat{\bf x} \right)\cdot \nabla \Psi - v_y \partial_y \Psi+\frac{c^2}{4\pi \sigma} \nabla^2 \Psi\,.
\eeq
\begin{align}
\partial_t B_{x}=&-\frac{c}{4 \pi e}\Bigg[\left(\nabla \left(\frac{\nabla^2 \Psi}{n_{e}}\times  \hat{\bf x}  \right)\cdot \nabla \Psi \right) \nonumber \\
&\ \ \ \ \ \ \ \ \ \ \ + B_x \left(\nabla n_{e}^{-1} \times  \hat{\bf x}  \right)\cdot \nabla B_x \Bigg] \nonumber \\
&+\left( \partial_z \Psi \partial_y v_{x}-\partial_{y}\Psi \partial_z v_{x} -v_{y} \partial_y B_{x}\right)\nonumber \\
&+\frac{c^2}{4\pi \sigma} \left(\nabla^2 B_{x} -\sigma^{-1}\nabla B_y \cdot \nabla \sigma\right)\,.
\label{EMHD2D}
\end{align}
We apply periodic boundary conditions at $y=0$ and $y= R_{nd}-R_{cc}$, which are implemented by a series of ghost-points on either side of the boundary. At the crust-core boundary we apply the zero boundary condition, therefore $\Psi(y, 0)=B_x(y,0)=0$. Finally, beyond $z=R_{nd}$ we consider the magnetic field to be current-free. Thus, $B_y(y,z>R_{nd})=0$, and further demanding that the Laplacian of $\Psi$  vanishes: $\Psi_{yy}(y,z>R_{nd})+\Psi_{zz}(y,z>R_{nd})=0$, which is implemented through two rows of ghost points.

A plastic flow will appear only if the crust fails. We scan through the crust checking whether the von Mises criterion is satisfied; if this is the case we integrate Eq.~\eqref{stokes_new}. This velocity is then included in the numerical integration of Eq.~\eqref{EMHD2D}. We note that the $y-$component has only a $z-$dependence because of its zero-divergence, whereas the right-hand-side of Eq.~\eqref{stokes_new} in principle depends on both $(y,z)$. To reconcile this issue we take the $y-$average of Eq.~\eqref{stokes_new}, which leads to an equation where both sides depend on $z$ only. 

Numerically solving for $\vpl$ at every timestep would drastically increase the computational time required to evolve the system. For this purpose, we repeat the survey of crust failure and solution for $\vpl$ not at every integration time-step, but less frequently. For the typical resolution employed here ($100^2$), we found that updating $\vpl$ every $10^3$ steps gives the same results as solving for $\vpl$ at every time-step, and hence we have adopted this for our calculation.

\subsection{Numerical tests}

We have run the code experimenting with various initial conditions and resolutions. For all initial conditions, we have run simulations where the field evolves only via electron MHD. These simulations exhibit the basic characteristics of such studies: the initial evolution is drastic, essentially responding to the non-equilibrium starting state. Following this initial phase, the system later relaxes to a Hall-attractor.  The pure electron MHD state serves as our benchmark simulation for a crust that does not yield and will be used for comparison with the runs with plastic flow. We have repeated some runs at higher resolution ($200^2$) and we found that the overall evolution is in good agreement with the integration times we present in this work. In the calculation we have included a finite resistivity $\sigma$ which leads to long-term decay of the magnetic field and an overall decrease in the magnetic energy of the field. \skl{The relative importance of Hall and Ohmic terms in the evolution may be determined from the magnetic Reynolds number (sometimes known as the Hall parameter):}
\beq
R_M = \frac{\sigma |B|}{c e n_{e}}\,.
\eeq
While $R_M$ is not constant throughout the crust, in the range of parameters employed we have a maximum value in our simulations in the order of few times $10^{2}$. We remark that in the setups with the strongest magnetic fields the maximum value was close to $10^{3}$ and the simulations were becoming numerically unstable. There we adopted a resistivity four times our standard value, to ensure numerical stability and to give a manageable $R_M\sim 10^2$.

\subsection{Magnetic-field reference state}

The magnetic field must reach an equilibrium state whilst the star is still fluid, so that when the crust first freezes it does so with the field in a particular state. It is born relaxed, naturally. Later it will deviate from this and so stresses build up. In studies of electron MHD without crustal failure the late-time evolution of the crustal field was somewhat independent of the initial state, with these early features having been washed out over time. One novel feature about the problem we study here is that the initial-state magnetic field remains important long after the crustal field has evolved, because it continues to set the crust's relaxed state. In other words, in principle one could remove all elastic stresses and the crust would return to this state (or vice-versa). We do not anticipate that plastic flow will allow significant adjustment of the reference state, since it does not act to relieve all crustal stresses -- only those exceeding the yield limit.

The initial conditions we implement are motivated by the presence of a strong crustal field, which appears in the form of loops resembling the structure of a poloidal loop field. We primarily consider an initial configuration with a double loop of closed field lines: 
\begin{align}
\Psi_{DC}(y,z)=B_0 \frac{[(z-R_{nd})(z-R_{cc})(x-R_{nd})(x-R_{cc})]^2}{\Delta R^7}\nonumber \\
\times \left[1+\sin\left(\frac{4\pi (x-R_{cc})}{\Delta R}\right)\right]\,,
\label{Ref1}
\end{align}
\begin{align}
B_{xDC}(y,z)=0\,.
\end{align}
Here $B_0$ is a parameter that sets the magnetic field strength, which is chosen so that the maximum value of the magnetic field is $\sim10^{15}$G. 

To verify that the conclusions we draw from our evolutions are general, and not peculiar to this choice of initial magnetic field, we will also try other initial conditions. One of these is a single-loop configuration:
\begin{align}
\Psi_{SC}(y,z)=B_0\frac{(z-R_{nd})(z-R_{cc})[(x-R_{nd})(x-R_{cc})]^2}{\Delta R^5}\,
\end{align}
\begin{align}
B_{xDC}(y,z)=0\, ,
\end{align}
although this state, like the previous double-loop configuration, features a magnetic field that vanishes on the surface. We will, therefore, also relax this assumption by allowing for some of the initial field lines to extend beyond $R_{nd}$, both in the double loop
\begin{align}
\Psi_{DO}(y,z)=B_0 \frac{[(z-R_{nd})(z-1.001R_{cc})(x-R_{nd})(x-R_{cc})]^2}{\Delta R^7}\nonumber \\
\times \left[1+\sin\left(\frac{4\pi (x-R_{cc})}{\Delta R}\right)\right]\,,
\end{align}
\begin{align}
B_{x DO}(y,z)=0\,.
\end{align}
and the single loop configuration:
\begin{align}
\Psi_{SO}(y,z)=B_0\frac{(z-R_{nd})(z-1.001R_{cc})[(x-R_{nd})(x-R_{cc})]^2}{\Delta R^5}\,,
\end{align}
\begin{align}
B_{xDO}(y,z)=0\,.
\label{Ref6}
\end{align}
In the non-dissipative problem, these initial conditions would set the reference state. However, in some models the dissipation is higher than what it would be for a realistic neutron star. This is primarily because we simulate a slab instead of the whole crust, thus large-scale configurations which should be long-lived are confined to the size of the simulation box -- and this sets the length-scale of the system and leads to faster dissipation. Secondly, in some case a higher resistivity is used than the realistic approximation. This could cause considerable energy loss, i.e.~a dissipation of half of the magnetic energy within 1 kyr. Thus the initial conditions do not represent a state which can ever be attained again by the field, since there is simply not enough magnetic flux left. Numerically, we have found that a permanent stress -- whose location and magnitude do not evolve significantly after a while -- develops independently of the details of the evolution. For this reason it is necessary to update the reference state, so that it represents a configuration with the same energy as the present-day one, with just the geometry being different. We do this by rescaling the $\maxw_0$, multiplying it by $E(t)/E(0)$ where $E(t)$ is the total magnetic energy of the simulated domain at time $t$ and $E(0)$ is its initial value. This is better than pumping up the evolving magnetic field because of possible numerical instabilities. The only thing to bear in mind is that whilst the whole magneto-plastic evolution is made self-consistent in this way, the characteristic timescale grows due to the reduction of $B$. So, the evolutions we see slow down at later times, despite the physical time moving forward at the same rate.

%%%%%%%%%%%%%%%%%%%%%%%%%%%%%%%%%%%%%%%%%%%%%%%%%%%%%%%%%%%%
%%%%%%%%%%%%%%%%%%%%%%%%%%%%%%%%%%%%%%%%%%%%%%%%%%%%%%%%%%%%
%%%%%%%%%%%%%%%%%%%%%%%%%%%%%%%%%%%%%%%%%%%%%%%%%%%%%%%%%%%%
\section{Results}

As the parameter space of our magneto-plastic evolution models is rather large we will explore some characteristic cases to illustrate possible avenues for the magnetic field evolution through electron MHD and plastic flow. In our study we focus on the following main parameters: magnetic field strength and structure, magnitude of the plastic viscosity $\nu$, failure criterion and evolution of the reference state. The simulations we have performed are detailed in table \ref{TAB1}.

For all models with a given magnetic field strength and structure we run a `Hall' simulation not permitting any plastic flow (H0-H6), a `plastic' simulation with plastic flow and where the reference state is rescaled (P0-P9) with the overall magnetic field decay, and a `non-rescaled' plastic simulation (N0-N9).
We have varied the maximum strength of the magnetic field and its structure; see the simulations with numbers 0 through 6.
We take $\nu_{max}=1.6\times 10^{38}$g~cm$^{-1}$~s$^{-1}$ as our standard value for the viscosity of the crust's plastic phase, but given the large uncertainties in the physical properties of the crust we have also experimented by decreasing its value by one or two orders of magnitude, to $\nu_{max}\sim 10^{37}$g~cm$^{-1}$~s$^{-1}$ (runs P7 and N7) and $\nu_{max}\sim 10^{36}$g~cm$^{-1}$~s$^{-1}$ (runs P8 and N8). \skl{(From now on, we suppress mention of the clumsy $1.6$ prefactor on the maximum plastic viscosity.)} We have also explored the possibility of collective failure occurring only once the von Mises criterion is satisfied deep within the crust, i.e. in the inner half of the simulated crust slab $z<(R_{cc}+R_{nd})/2$ (runs P9, N9), as opposed to our standard assumption that failure occurs when any gridpoint satisfies the failure criterion -- effectively meaning a shallow failure.

We simulate all models for a timescale of $5$ kyrs to allow for sufficient time for the system to evolve fully. We restrict the timescale to this level, as beyond this timescale the evolution practically saturates to a Hall-attractor \citep{Gourgouliatos:2014} state. 

\begin{table*}
\begin{center}
\caption{Summary of the simulations, all of which have a total run time of 5 kyr. The first column is the run name, and subsequent columns are: the maximum value of magnetic field occurring in the crust; the reference state described by equations \eqref{Ref1}-\eqref{Ref6} (where $SC,DC$ refer to Single Closed loop and Double Closed loop, and $SO,DO$ the corresponding states with open field lines at the surface); the maximum value of the plastic viscosity; the maximum value of conductivity used; whether plastic flow is permitted (Y) or not (N); whether we allow for plastic flow as soon as the failure criterion is satisfied \emph{anywhere} -- effectively meaning a shallow failure (S) -- or whether we wait for it to be satisfied in the deeper part of the crust (D); and whether the reference state is rescaled (Y) or not (N). }
\begin{tabular}{ |c|c|c|c|c|c|c|c|c|c}
\hline
 Model & $B_{\rm max}$ & RS & $\nu_{\rm max}$ & $\sigma$ & Plastic&  Failure & Rescaling \\
  & ($10^{15}$G)& & ($1.6\ \textrm{g\ cm}^{-1}\textrm{s}^{-1}$)&(s$^{-1}$) & & & \\
 \hline
 H0 & 1 & DC & -& $10^{24}$  &N& -& -\\ 
 N0 & 1 & DC &$10^{38}$ & $10^{24}$  & Y& S & N  \\ 
 P0 & 1 & DC &$10^{38}$ & $10^{24}$  & Y& S & Y  \\
\hline
 H1   & 1.3 &DC &- & $10^{24}$  & N& -& - \\ 
 P1 & 1.3 & DC &$10^{38}$ &$10^{24}$  & Y& S & Y \\ 
 N1 & 1.3 & DC &$10^{38}$ &$10^{24}$  &  Y&S & N\\
  \hline
  H2 &2&DC&   - &$10^{24}$  & N& - & - \\ 
 P2 & 2 &DC &$10^{38}$ & $10^{24}$  &  Y&S & Y \\ 
 N2 & 2 & DC &$10^{38}$ &$10^{24}$  &  Y& S & N\\
  \hline
  H3 & 2 & DO &-& $10^{24}$  & N& - & -  \\ 
 P3 & 2 & DO &$10^{38}$ &$10^{24}$  & Y& S & Y \\ 
 N3 & 2 & DO &$10^{38}$ &$10^{24}$ & Y& S & N\\
  \hline
    H4 &4& SC &- & $10^{24}$  & N& - & - \\ 
 P4& 4& SC &$10^{38}$ &$10^{24}$  & Y& S & Y \\ 
 N4 & 4 & SC &$10^{38}$ &$10^{24}$ & Y& S & N\\
  \hline
    H5 &4& SO&- &  $10^{24}$ &  N& -& - \\ 
 P5 & 4 &SO & $10^{38}$ &$10^{24}$   & Y&S & Y \\ 
 N5 & 4 & SO & $10^{38}$ &$10^{24}$   & Y& S & N\\
  \hline
    H6 &4& DC &- &  $2.5\times 10^{23}$ & N &- & - \\ 
 P6 & 4& DC &$10^{38}$ &$2.5\times 10^{23}$ &Y& S &Y \\ 
 N6 & 4 & DC &$10^{38}$ &$2.5\times 10^{23}$ &Y& S & N\\
 \hline
 P7& 2& DC &$10^{37}$ & $10^{24}$ &Y& S &Y   \\ 
 N7 & 2& DC &$10^{37}$ &$10^{24}$ &Y & S & N\\
 \hline
 P8  & 2&DC &$10^{36}$ & $10^{24}$ & Y& S &Y \\ 
 N8 & 2& DC &$10^{36}$ &$10^{24}$  & Y & S &N \\
  \hline
 P9 &2&DC &$10^{38}$ & $10^{24}$  & Y & D & Y  \\ 
 N9 & 2 &DC &$10^{38}$ & $10^{24}$  & Y & D &N \\
\hline
\end{tabular}
\label{TAB1}
\end{center}
\end{table*}

\subsection{Crust failure}

\begin{figure*}
\begin{center}
\begin{minipage}[c]{0.7\linewidth}
a\includegraphics[width=0.5\textwidth]{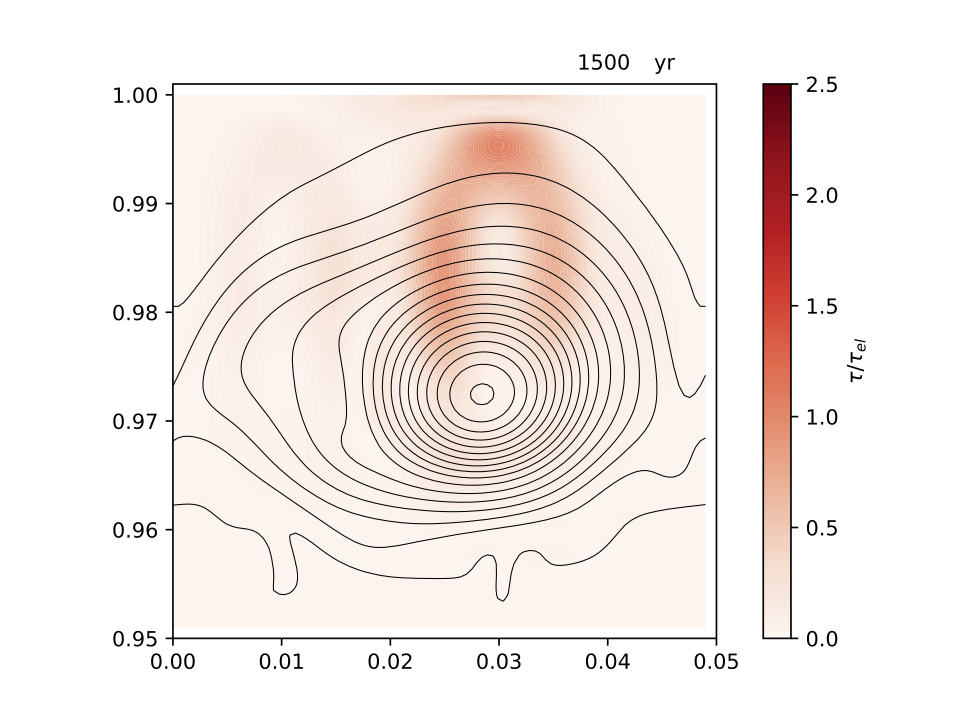}
b\includegraphics[width=0.5\textwidth]{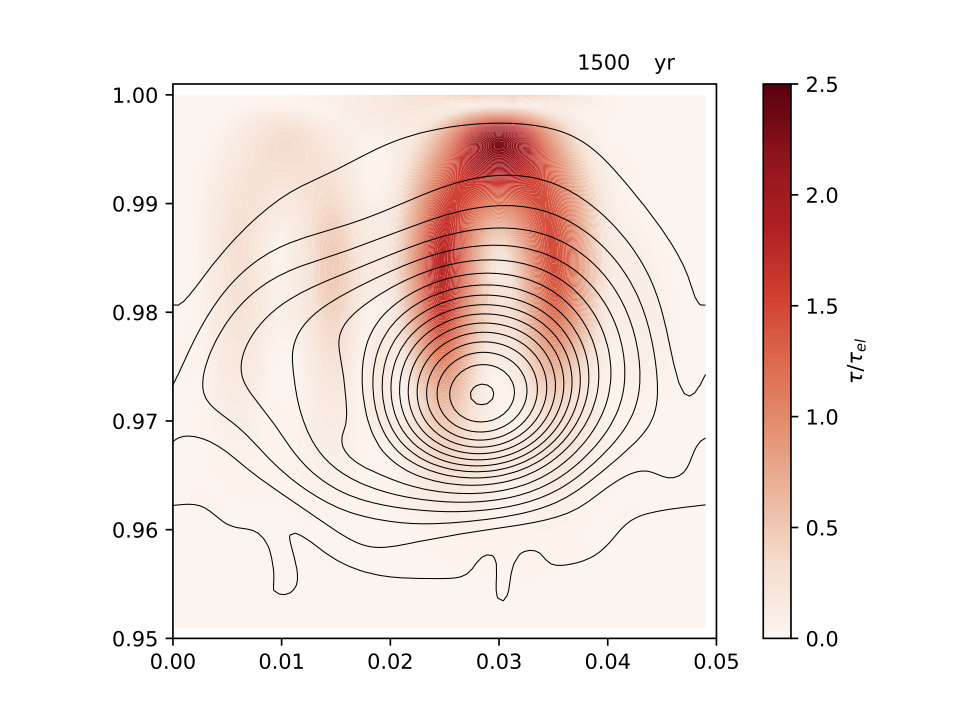}
\end{minipage}
\caption{\label{Panel1_5}
Black contours indicate the poloidal ($B_y$--$B_z$) structure of the field, and the ratio of $\tau/\tau_{el}$ is shown in colour. Panel (a) corresponds to simulation P2 and panel (b) to simulation N2, both at $t=1500$ years. \skl{This demonstrates the more physically realistic nature of the P-simulations, which account for the decreasing total magnetic energy of the crust reference state, and display a plastic flow that keeps stresses relatively low. In the N-simulations, by contrast, a large and permanent stress is seen. }}
\end{center}
\end{figure*}

We find that for the weakest magnetic field strength simulated (H0, P0, N0), the von Mises criterion is never satisfied, thus no failure occurred during the simulation. An interesting point to notice for this system is that even if the magnetic field were to be completely removed by the evolution, the von Mises criterion would still barely be satisfied.

Setups with stronger initial magnetic field magnetic field (H1, P1, N1) demonstrate the importance of applying the rescaling to the magnetic field reference state, to compensate for the decay of the magnetic field. In run P1 the von Mises criterion is not satisfied and the system does not initiate a plastic flow. On the contrary, in run N1, where the reference state is not rescaled, a plastic flow is initiated after $300$ years. This is mainly due to the fact that approximately $15\%$ of the magnetic field energy has dissipated and consequently the von Mises criterion is more easily satisfied. Similarly, in simulation P2, stresses hardly exceeds $\tau_{el}$ anywhere in the domain, whereas in simulation N2 where no rescaling is applied the stresses are locally $2.5$ times higher than $\tau_{el}$. To show the effect of the rescaling on the evolution, we compare models with and without this effect in Fig.~\ref{Panel1_5}, finding that plastic flow is more effective in reducing stresses beyond the von-Mises limit when rescaling is included -- a physically more reasonable result than the stresses persisting indefinitely.

The family of simulations consisting of a double loop confined inside the star where the maximum magnetic field is $2\times 10^{15}$~G (P2-N2, P7-N7, P8-N8) initiate a plastic flow within the first 40 years. By constrast, simulations with a double loop where the initial state has open field lines develop a plastic flow much earlier at around 10 years (P3-N3). Simulations with a single loop reach the plastic regime more slowly: at around 80 years if the initial flux is contained within the crust (P4-N4) and around 50 years if there are also open field lines (P5-N5). We \skl{can understand this by noting} that while the single loop initial conditions have the same maximum value of the magnetic field the electric currents are weaker, due to the simpler structure of the initial field and thus the overall stresses. Similarly, the models where the initial magnetic field has field lines emerging from the surface of the crust, yield earlier due to the stronger initial field near the surface. The simulations P9 and N9, where we demand the von Mises criterion to be satisfied in the inner half of the crust before allowing for failure (i.e. plastic flow), take 80 years to fail -- twice as long as the same family of simulations where the von Mises criterion need only be satisfied at shallow depths (P2-N2).

Finally, we note that the simulations where the reference state is not rescaled will continuously fulfil the von Mises  criterion once they fail, since they cannot get back to the initial conditions because of the magnetic energy that has since been lost; this is the case for all N-simulations except for the N0 run. On the contrary, some of the simulations where the reference state is rescaled (P2, P3, P7, P8 and P9) tend after some time, ranging from 500-2500 years, to stop fulfilling the von Mises criterion (Eq.~\ref{vonmises_mag}). This is primarily due to the fact that the maximum stress they developed was only a factor of about 1.5 that of the (evolving) $\tau_{el}$, and it became possible for the plastic flow to return the {magnetic-field structure to a state where its stresses could be fully absorbed by the crust again.} On the contrary, in simulations P4, P5 and P6, the von Mises criterion is satisfied with the right-hand-side being a factor of 2-5 higher than $\tau_{el}$. In this case, the magneto-plastic evolution cannot push the system sufficiently close to the reference state to choke the plastic flow.

\subsection{Plastic flow}

\begin{figure*}
\begin{center}
\begin{minipage}[c]{\linewidth}
a\includegraphics[width=0.32\textwidth]{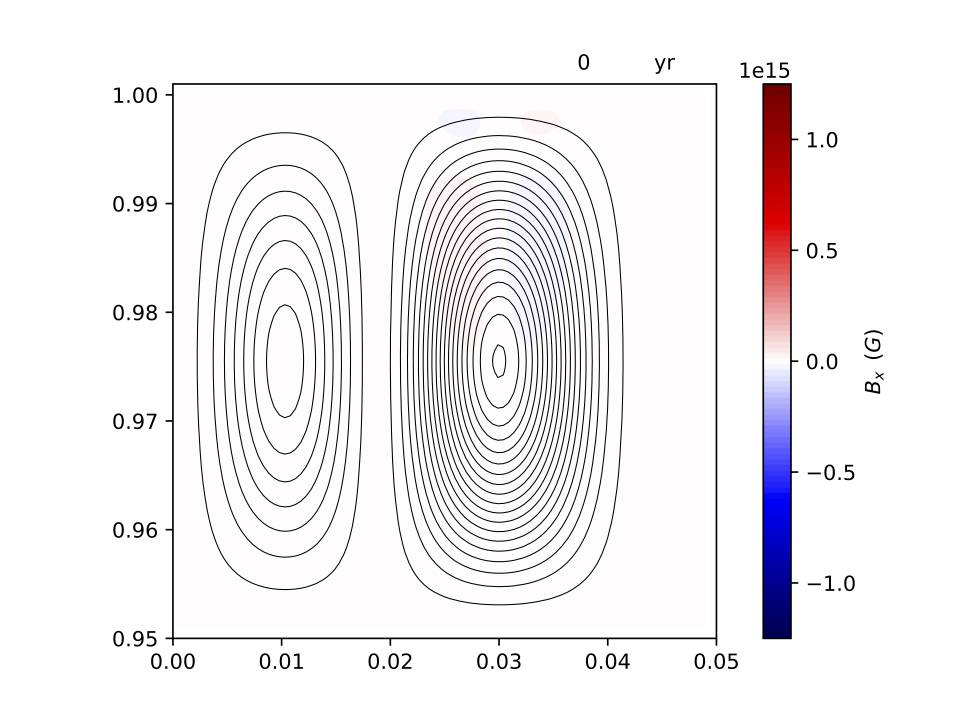}
b\includegraphics[width=0.32\textwidth]{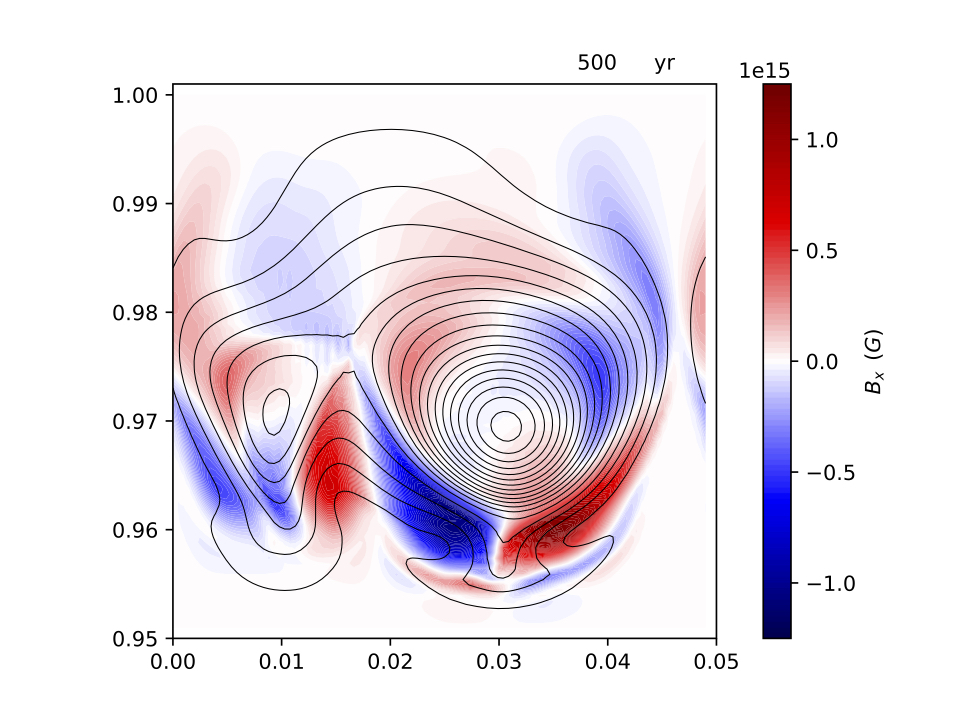}
c\includegraphics[width=0.32\textwidth]{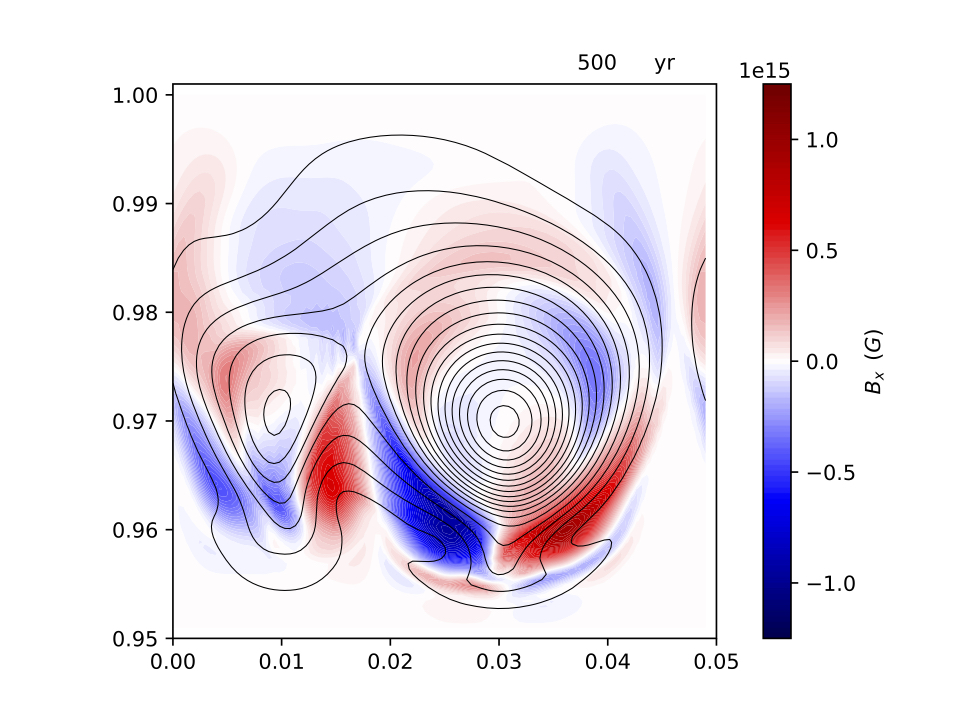}
d\includegraphics[width=0.32\textwidth]{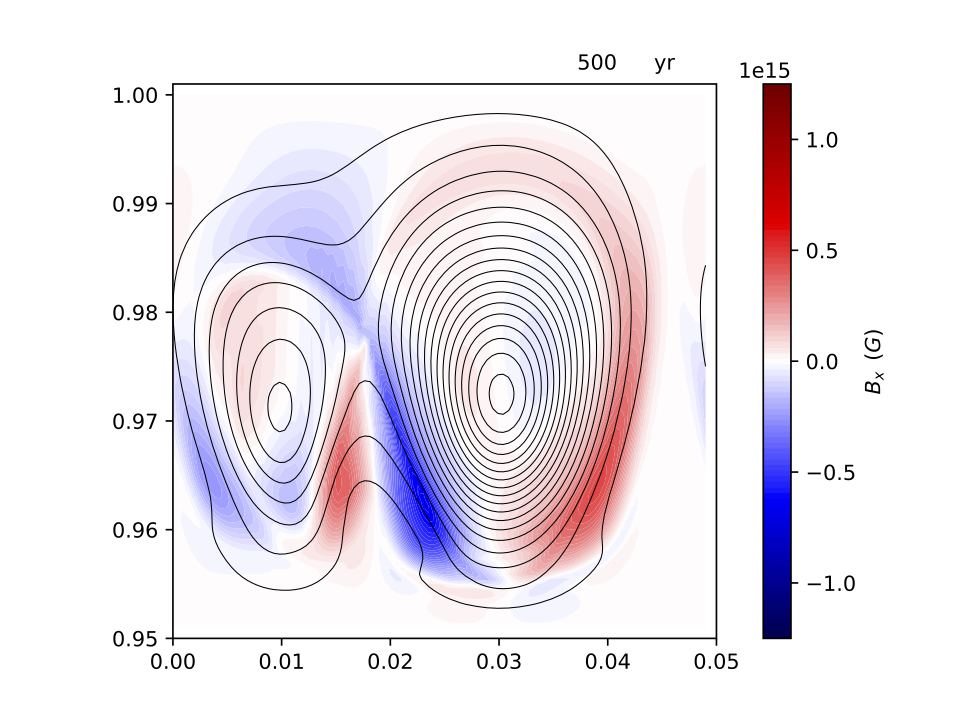}
e\includegraphics[width=0.32\textwidth]{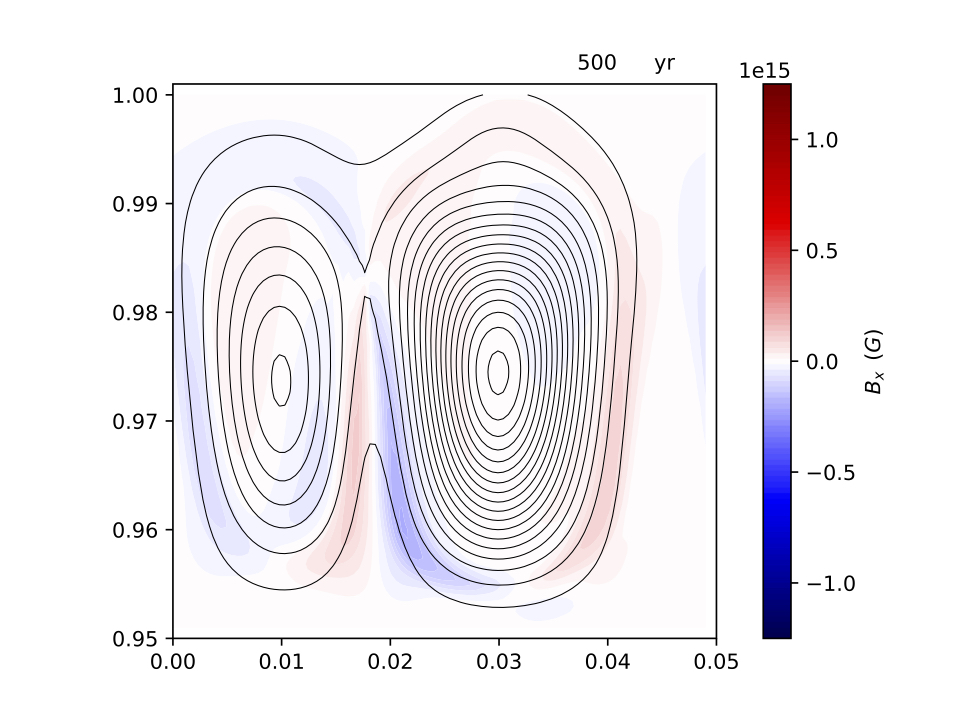}
f\includegraphics[width=0.32\textwidth]{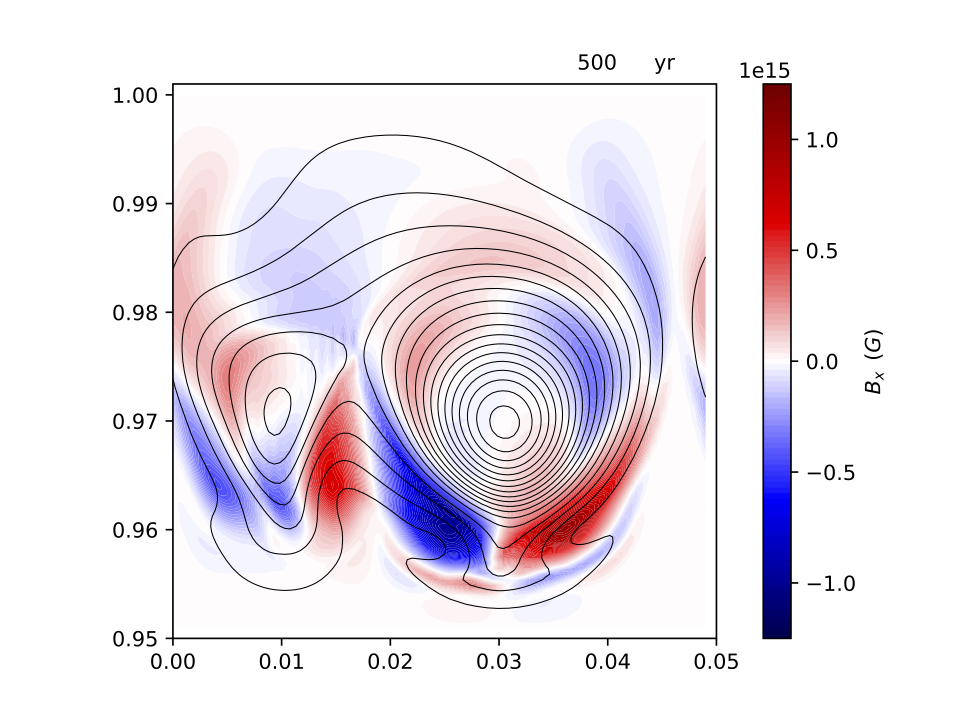}
\end{minipage}
\caption{\label{Panel1}
Black contours indicate the poloidal ($B_y$--$B_z$) structure of the field, and the intensity of the toroidal component $B_{x}$ is shown in colour. \skl{The same colourscale is used for all panels, for ease of comparison.} Panel (a) is the initial state at $t=0$ for all runs while the rest of the panels show the structure of the field at $t=500$ years, with (b) no plastic flow (H2), (c) high-viscosity plastic flow (P2), (d) intermediate-viscosity plastic flow (P7), (e) low-viscosity plastic flow (P8), (f) high-viscosity plastic flow with deep failure criterion (P9).}
\end{center}
\end{figure*}
\begin{figure*}
\begin{center}
\begin{minipage}[c]{0.7\linewidth}
a\includegraphics[width=0.5\textwidth]{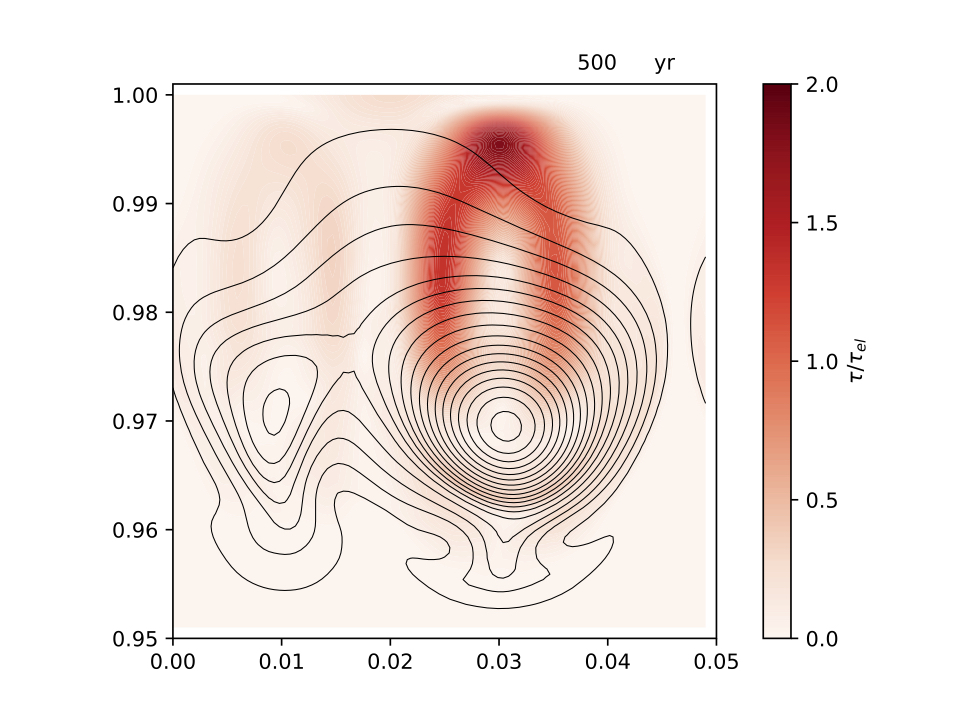}
b\includegraphics[width=0.5\textwidth]{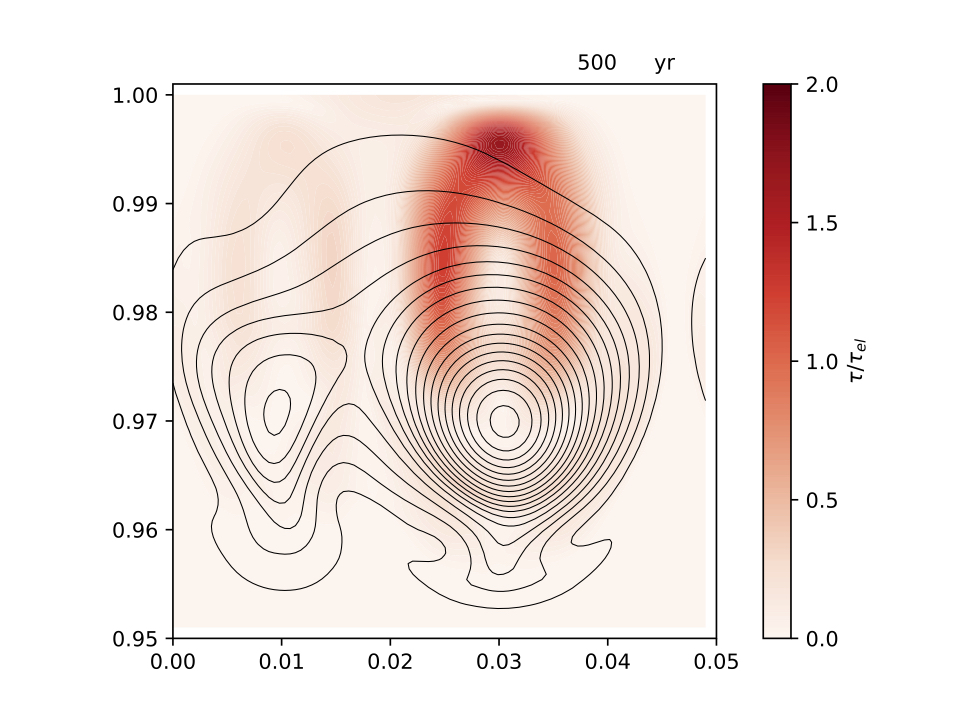}
c\includegraphics[width=0.5\textwidth]{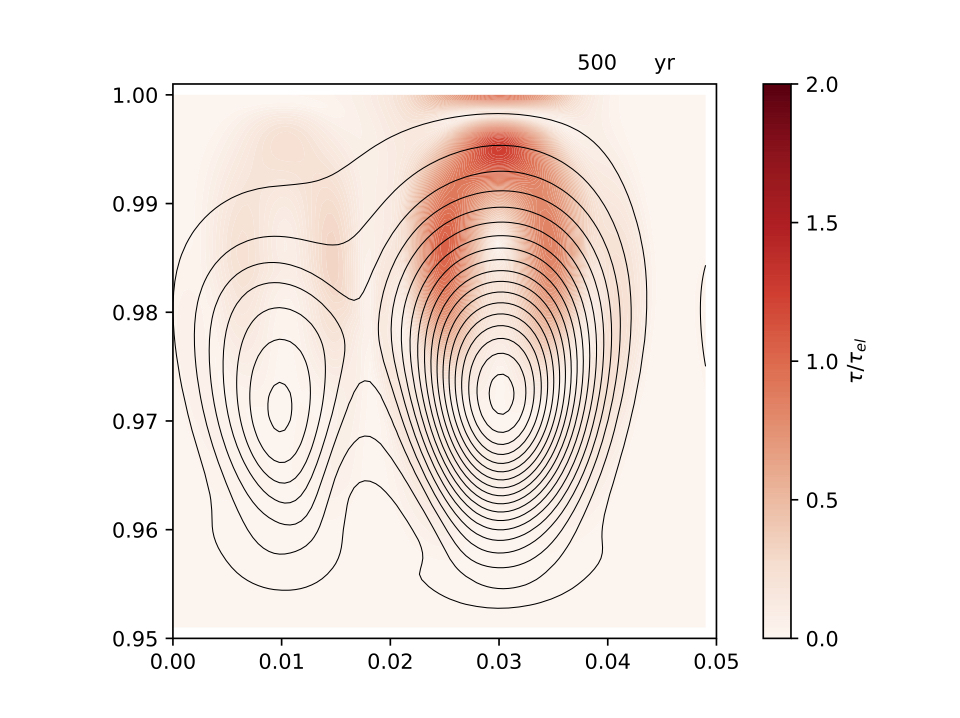}
d\includegraphics[width=0.5\textwidth]{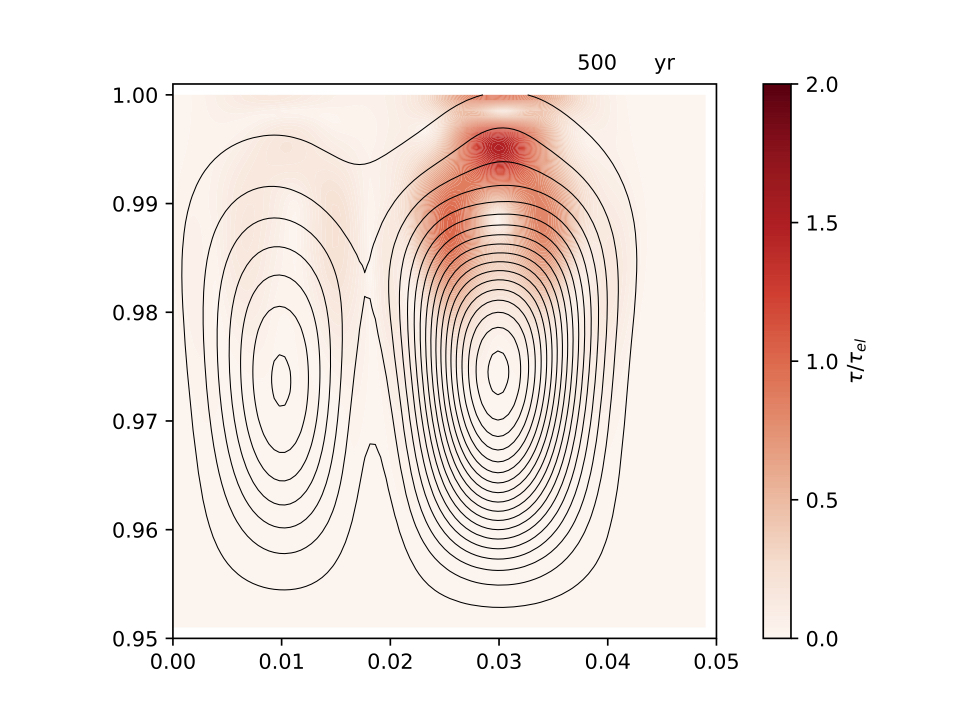}
\end{minipage}
\caption{\label{Panel2}
Colourscale shows the ratio of $\tau/\tau_{el}$, and black contours again show the poloidal field lines, at time $t=500$ years. The same colourscale is used for all panels. The panels correspond to runs (a) no plastic flow (H2); (b),(c),(d) high-, intermediate- and low-viscosity plastic flow (P2,P7,P8).  }
\end{center}
\end{figure*}

In all simulations where a plastic flow develops we find that it tends to slow the pure electron MHD evolution. In particular, the plastic velocity that develops tends to oppose the electron flow velocity, whenever this is possible -- as $\vpl$ is subject to more constraints than $\bv_e$. In all simulations we have deliberately set $B_x(t=0)=0$, thus this component can be thought of as a proxy for the efficiency of electron MHD evolution: the initial electron velocity distribution winds up the poloidal $B_y$--$B_z$ components of the field and develops a toroidal $B_x$ field. If the plastic flow is in the opposite direction from the electron flow, the $B_x$ component that will develop will be weaker than pure electron MHD.

In Fig.~\ref{Panel1} we compare the state of the magnetic field at $t=500$ yr in simulations H2, P2, P7, P8 and P9. Since all these have identical initial conditions we can deduce the impact of the plastic flow and the role of the value of plastic viscosity. The system that evolves entirely via electron MHD develops a maximum for $B_x$ of about $1.2\times 10^{15}$~G (Fig.~\ref{Panel1} b). If a plastic flow is permitted, then the maximum value of the field drops, with the drop being dramatic for decreasing plastic viscosity: for $\nu\sim 10^{38} \textrm{g\ cm}^{-1}\textrm{s}^{-1}$ the maximum value is $10^{15}$~G (Fig.~\ref{Panel1} c), whilst decreasing this parameter by one and two orders of magnitude results in the maximum field strength dropping to $6.5 \times 10^{14}$~G (Fig.~\ref{Panel1} d) and $2\times 10^{14}$~G (Fig.~\ref{Panel1} e) respectively. Allowing for plastic flow only when the inner part of the crust fails (Fig.~\ref{Panel1} f) gives only a marginal difference from the case where the failure criterion is applied globally. In all cases we notice that plastic flow leads to smoother magnetic field lines and consequently weaker electric currents. 

As expected, the plastic flow relieves part of the stress on the crust, as can be seen by comparing the normalised stresses $\tau/\tau_{\rm el}$ of models H2, P2 , P7 and P8 -- see Fig.~\ref{Panel2}. Note that it is not always the system with the lowest $\nu$ that has least stress -- in these particular snapshots, it is the system with $\nu\sim 10^{37}  \textrm{g\ cm}^{-1}\textrm{s}^{-1}$. 

\begin{figure*}
\begin{center}
\begin{minipage}[c]{0.7\linewidth}
a\includegraphics[width=0.5\textwidth]{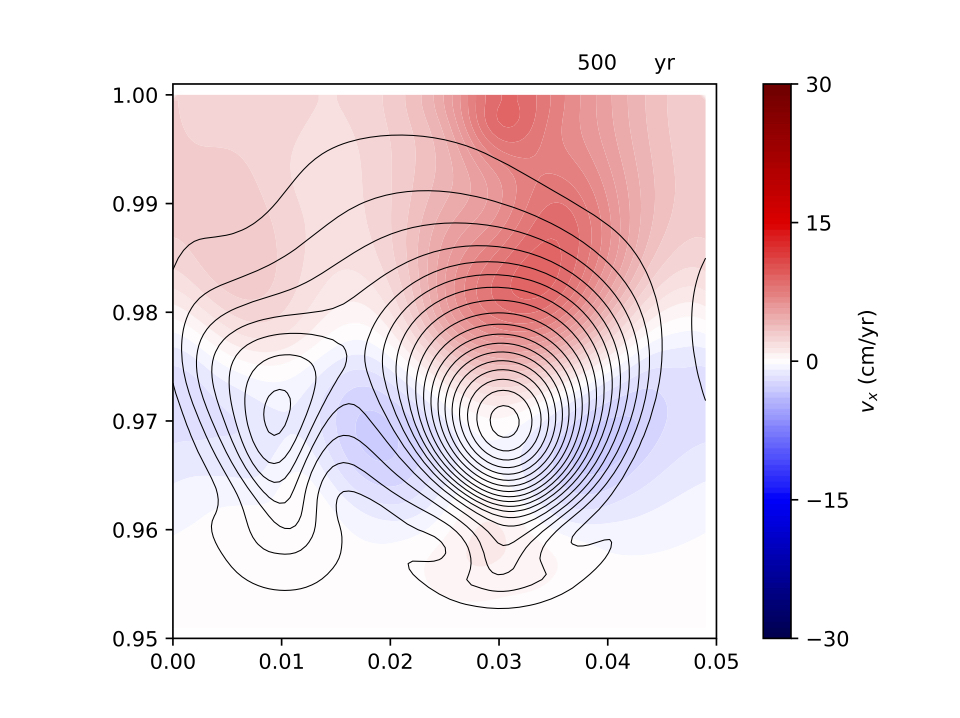}
b\includegraphics[width=0.5\textwidth]{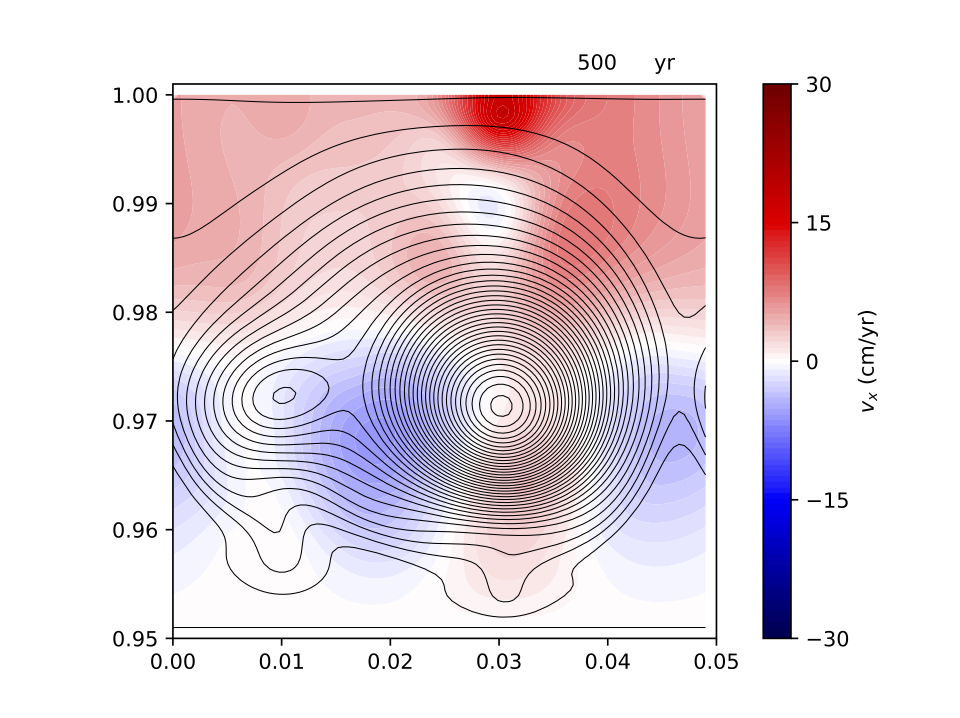}
c\includegraphics[width=0.5\textwidth]{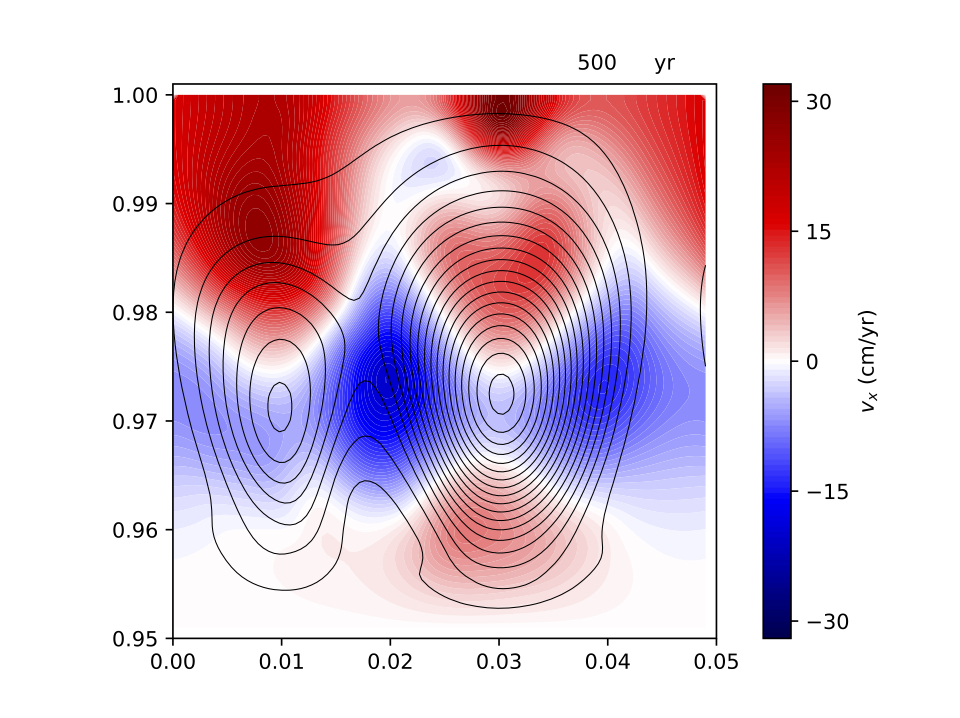}
d\includegraphics[width=0.5\textwidth]{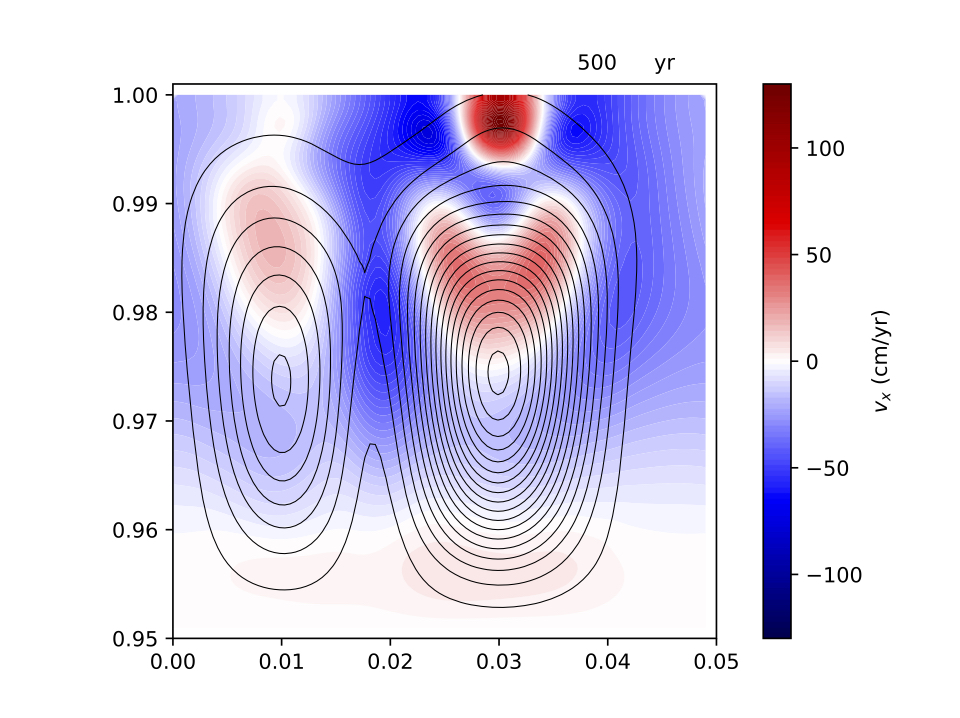}
\end{minipage}
\caption{\label{Panel3}
Colourscale shows the $x$ (azimuthal) component of the plastic flow velocity $\vpl$, against poloidal-field lines (black) as usual, at time $t=500$ years. In this case we do not use the same colourscale for all panels, as the velocities in panel (d) are considerably greater than for the other three cases. The panels correspond to runs (a) P2, (b) P6, (c) P7, (d) P8.}
\end{center}
\end{figure*}

\begin{figure*}
\begin{center}
\begin{minipage}[c]{0.7\linewidth}
a\includegraphics[width=0.5\textwidth]{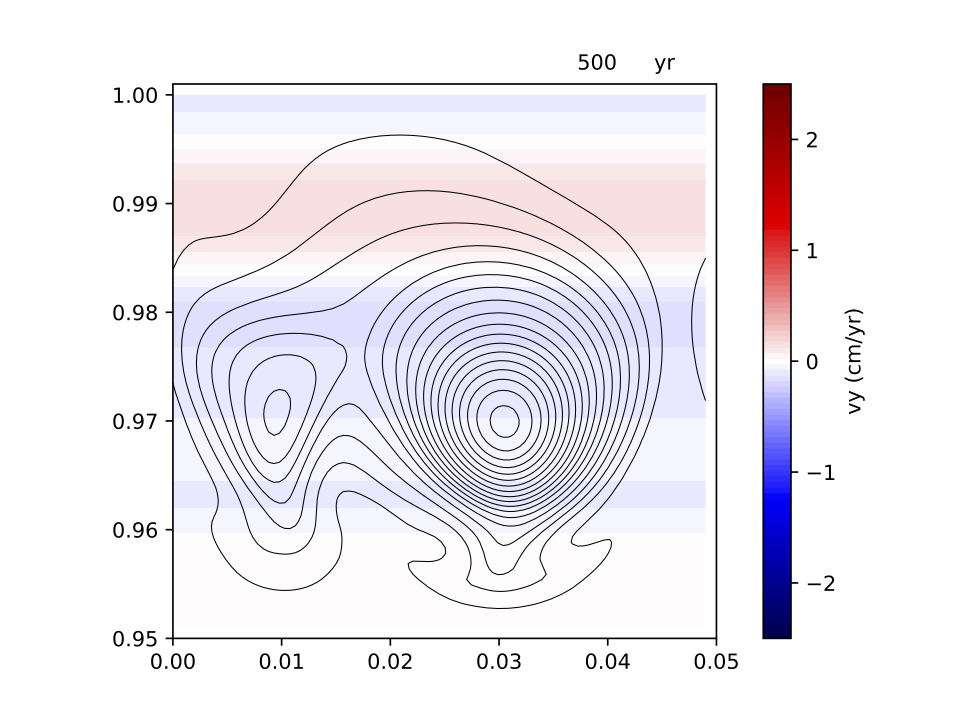}
b\includegraphics[width=0.5\textwidth]{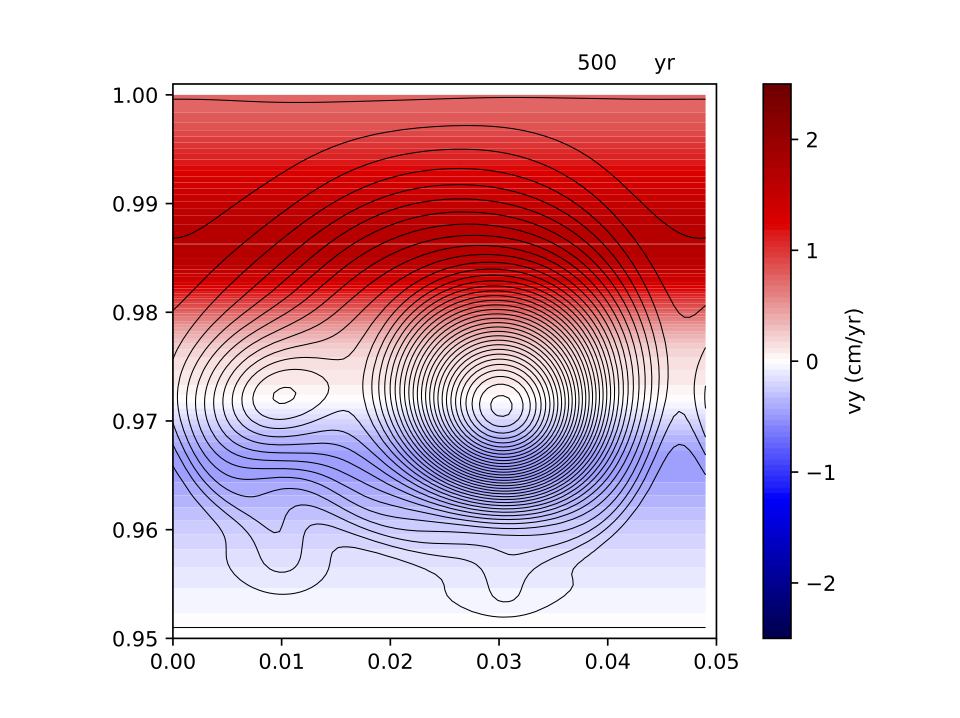}
c\includegraphics[width=0.5\textwidth]{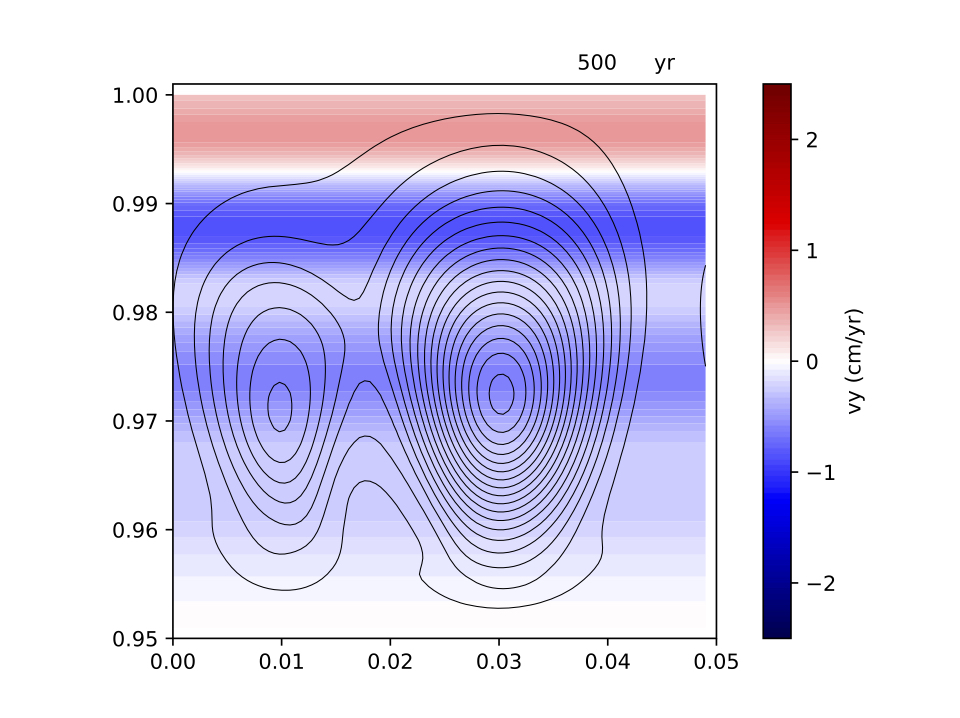}
d\includegraphics[width=0.5\textwidth]{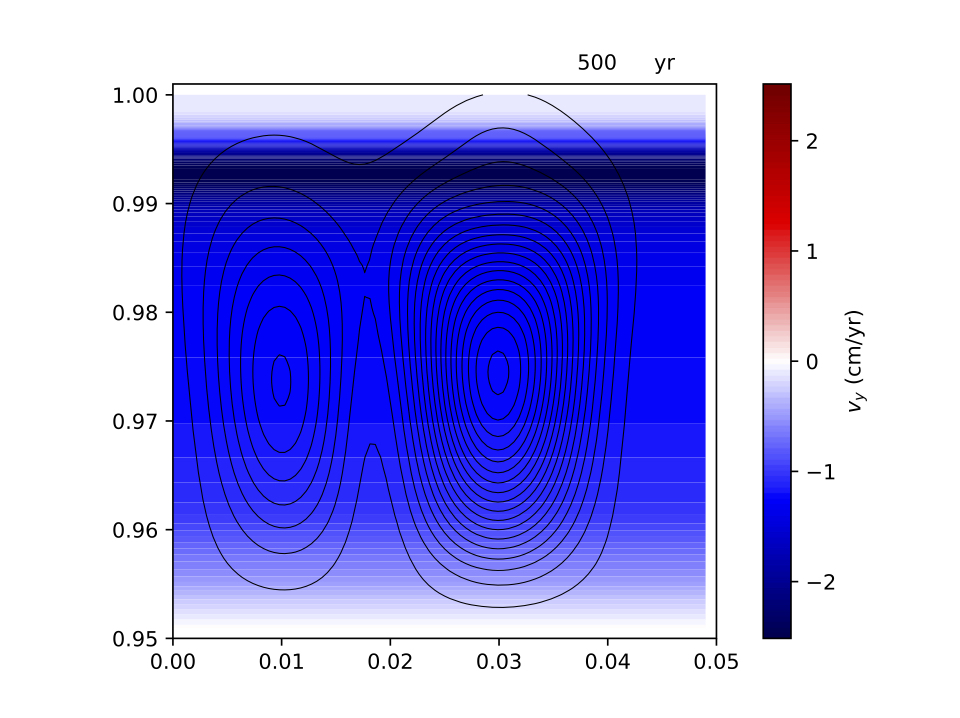}
\end{minipage}
\caption{\label{Panel4}
Colourscale shows the $y$ component of the plastic flow velocity $\vpl$, against poloidal-field lines (black) as usual, at time $t=500$ years. Recall that our plane-parallel geometry means we have to average this component so it depends on depth alone. The panels correspond to runs (a) P2, (b) P6, (c) P7, (d) P8.  }
\end{center}
\end{figure*}

The velocities of the plastic flow depend mainly on the plastic viscosity chosen and the initial magnetic field strength, Fig.~\ref{Panel3} and \ref{Panel4}. Comparing simulations P2, P6, P7 and P8 we notice that a drastically lower plastic viscosity leads to a higher velocity, e.g.~compare the maximum values of plastic velocities of $10$cm/year for P2 to $130$cm/year for P8. Similarly, stronger magnetic fields leads to faster plastic flows: doubling the magnetic field strength from simulation P2 to P6 we notice an increase by a factor of 3 on the plastic flow velocity. These maximum velocities occur at the upper layers of the crust, where the plastic viscosity is lower. Broadly speaking, from \skl{the snapshots of} Fig.~\ref{Panel3} we see a strong azimuthal plastic flow $v_{\rm pl}^x$ close to the surface, and a weaker internal flow in the opposite direction deep within the crust, \skl{though the detailed properties of the flow evolve in time}. The $v_{\rm pl}^y$ flow is simpler, an artefact of our need to perform an average for consistency with the plane-parallel geometry we have imposed, and is constant for a given depth within the crust. On a sphere it would represent a flow in the $\theta$-direction; within our local domain it simply flows from left to right, or vice-versa, with respect to the viewpoint of the plots of Fig. \ref{Panel4}.


\subsection{Energy dissipation}

We established in equation \eqref{vpl_conserv} that a magneto-plastic evolution must feature either non-zero radial plastic velocity or non-zero radial magnetic field at the surface in order to be conservative. The former quantity is always zero by assumption in our work, and the latter is also zero for the initial closed-loop configurations. However, a non-zero radial field develops by the time the plastic regime is entered, so we can expect the flow not to dissipate magnetic energy. The same is also true (without caveats) for Hall drift \citep{GR92}. Despite these two processes being conservative, they may still affect the overall magnetic field decay due to the Ohmic term, so we check this in Fig.~\ref{Panel4}. It is well known that the Hall effect can generate smaller-scale structures that are more readily dissipated by Ohmic decay, and so are indirectly responsible for magnetic-energy loss. The addition of a highly viscous plastic flow results in a mildly faster dissipation of magnetic energy, presumably numerical in origin, since the evolution is now more complex. Interestingly though, if a lower plastic viscosity is chosen (P8), the magnetic field energy decays slower than in the pure-Hall case. We believe this is because plastic flow acts to iron out some of the smaller-scale structures generated by Hall drift, and thus makes Ohmic decay less effective.
\begin{figure}
\begin{center}
\begin{minipage}[c]{\linewidth}
  %a
  \includegraphics[width=\textwidth]{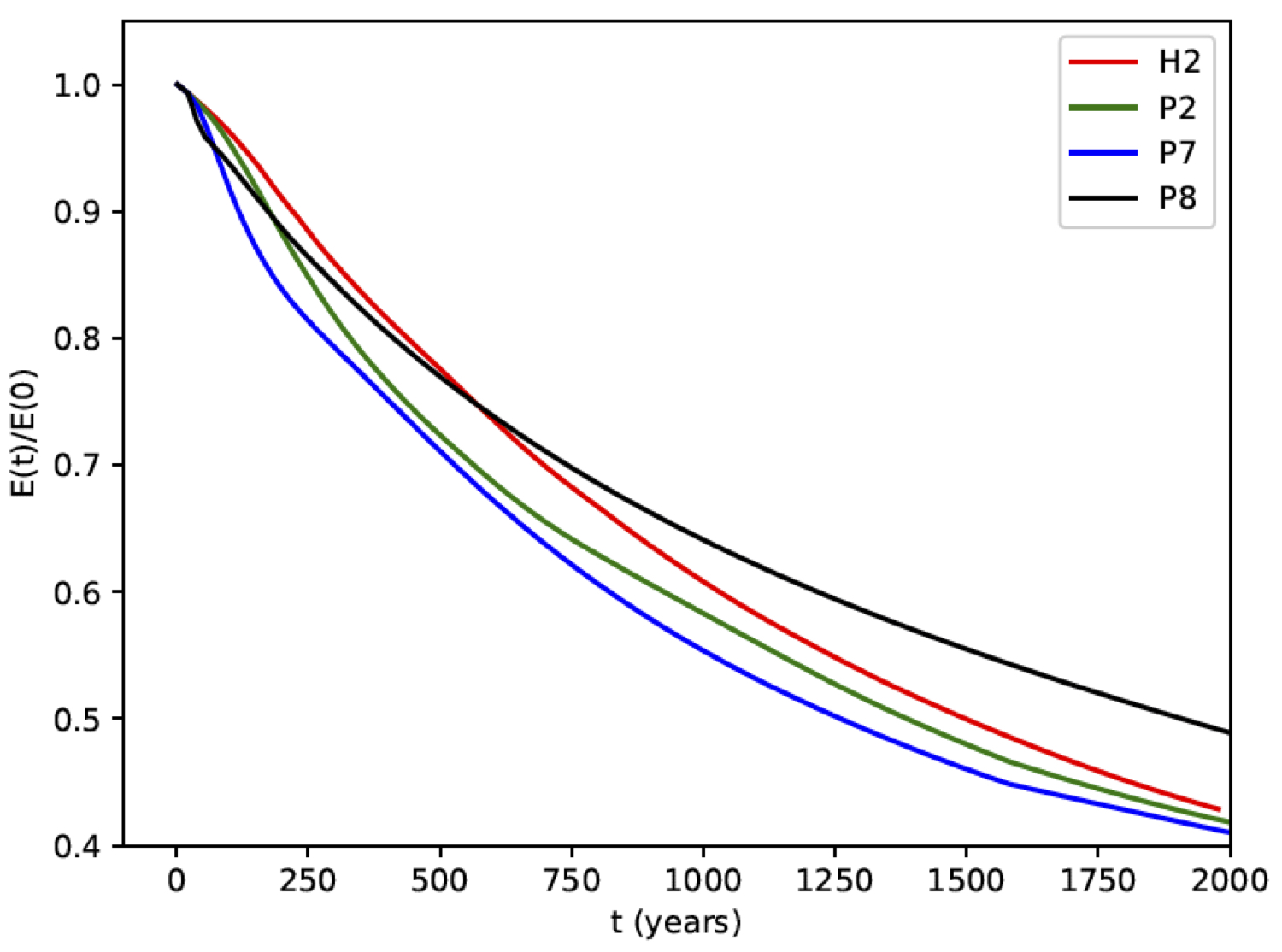}
%b\includegraphics[width=\textwidth]{figs/Panel5/Energy2.png}
\end{minipage}
\caption{\label{Panel5}
The energy stored in the magnetic field as a function of time, for simulations H2, P2, P7, P8.}
\end{center}
\end{figure}

%%%%%%%%%%%%%%%%%%%%%%%%%%%%%%%%%%%%%%%%%%%%%%%%%%%%%%%%%%%%
%%%%%%%%%%%%%%%%%%%%%%%%%%%%%%%%%%%%%%%%%%%%%%%%%%%%%%%%%%%%
%%%%%%%%%%%%%%%%%%%%%%%%%%%%%%%%%%%%%%%%%%%%%%%%%%%%%%%%%%%%
\section{Discussion}

In this paper we believe we have performed the most quantitative, self-consistent simulations to date of the magneto-plastic dynamics of a neutron star's crust beyond the elastic yield limit. Our main conclusions are the following:\\
(1) Magnetic fields in the inner crust that are below $10^{15}$ G everywhere barely lead to any plastic flow, as the deviations of the Maxwell stress tensor from its initial state (at which the crust was relaxed) can only ever be marginally above the yield stress.\\
(2) Broadly speaking, plastic flow opposes Hall evolution, slowing it down and limiting the development of sharp gradients in the magnetic field. A very low plastic viscosity practically chokes Hall evolution completely, and rather fast plastic velocities develop -- up to hundreds of centimetres per year.\\
(3) Increasing the magnetic-field strength, for a given (moderate) value of plastic viscosity does not completely halt the Hall effect, but merely slows it down. Since plastic velocities should scale like $B^2$, as opposed to the linear $B$-scaling for Hall velocities, an extrapolation of our results suggests that the evolution of magnetic fields between $5\times 10^{15}-10^{16}$ G will become dominated by the plastic-flow term.\\
(4) Once the crust fails, we find that the plastic flow could last for $500-2000$ years for our domain size, as long as we account for the global decrease in magnetic energy in calculating the crust's reference relaxed state. Afterwards it reaches some modified version of a `Hall attractor' state.\\

Our work has many interesting repercussions for the physics of highly-magnetised neutron star crusts, but before discussing these we need to be aware of various caveats. These are mostly related to the local nature of our simulations, where a small slab of crust is modelled under the assumption of plane-parallel geometry. Firstly, the effectiveness of plastic flow in counteracting Hall drift may be limited by the fact that  we place more assumptions on $\vpl$ than on $\bv_e$: in particular, we assume the former velocity to be incompressible, to have no $z$-component and to have a $y$-component which is obtained by an averaging procedure. Secondly, the small size of our domain means the characteristic lengthscale of the magnetic field is far shorter than the $1$-km scale one might expect for a global simulation. As a result all of our evolutions are considerably faster than would be expected for a full-crust evolution.

By looking at the results of our evolutions and whether they allow for the expected range of magnetar activity, we can finally place some physical bounds on the viscosity of the crust's plastic phase. Let us demand that surface motions are sufficiently fast to implant a substantial, persistent twist into the magnetosphere, as required by the corona model of \citet{belo_thomp,belo09}; this is only possible for a viscosity $\nu\lesssim 10^{37}\ \textrm{g cm}^{-1}\textrm{s}^{-1}$. Equally, features related to Hall drift tend to be washed out for $\nu\lesssim 10^{36}\ \textrm{g cm}^{-1}\textrm{s}^{-1}$ -- were the viscosity to be so low, we would be led to the undesirable need to dismiss all the literature associating features of magnetar evolution and emission with Hall drift. The noniconoclastic conclusion, avoiding conflict with previous work, is that $\nu\sim 10^{36}-10^{37}\ \textrm{g cm}^{-1}\textrm{s}^{-1}$.

We have shown on theoretical grounds, and from our simulations, that a magneto-plastic evolution generically results in substantial surface motions of the kind that would produce an azimuthal twist in the magnetosphere. This is a non-trivial result, as the flow could have been dominantly further into the crust or only weakly azimuthal. Dropping this assumption, we anticipate at least one qualitatively new feature: with $v_{\rm pl}^y$ less restricted, surface motions involving a circulation of plastically-deformed crustal matter. These will immediately render the magnetosphere non-axisymmetric, with a characteristic azimuthal structure dictated by the relative size of $v_{\rm pl}^x$ and $v_{\rm pl}^y$.

Our simulations, and indeed the consensus about neutron-star crustal failure, relies on the macroscopic plastic flow being a local effect, even though it is seen to be collective on microscopic scales \citep{CH10,hoff_heyl}. This allows for local twisting of magnetospheric field lines and the generation of a localised corona, and larger bursts or flares can be attributed to overtwisting and the resultant violent reconnection \citep{lyu03,parfrey,akgun18}. There is perhaps one dissenting voice to this. \citet{thomp17} consider a toy model for core-field evolution and based on its results argue that crustal failure is truly global: {by imposing a large-scale stress at the crust-core boundary, a localised yielding event in a small region can propagate across the whole crust. Their simulations show some interesting features that resemble observed magnetar activity, like the presence of lower-energy `aftershocks' after a giant flare. The model makes, however, some inevitable simplifications, which impede a quantitative comparison with neutron-star observables. In particular, their core field is not a hydromagnetic equilibrium solution, but rather modelled by an elastic spring, and their evolution -- changing the spring constant -- does not seem reflective of any known mechanism for core-field evolution.

We do, however, believe a viable picture for the giant flares of magnetars can be produced from the ideas of \citet{belo_lev} and \citet{thomp17}, and using some insight from our simulations. If the plastic flow were not strictly conservative but instead involved some viscous heating, the dynamics would be modified: excessive heating would probably prevent significant surface motion, but mild heating would reduce the local viscosity of the plastic phase and accelerate the evolution. A sustained episode of accelerated plastic flow would implant a magnetospheric twist growing on a timescale far shorter than that of exterior dissipation, and eventually might trigger an overtwisting instability \citep{mikic_link,parfrey}. Since the requisite balance between the plastic heating and flow rate may be rather delicate, it would not surprising to find that such energetic events are very rare. Certainly, some mechanism is needed to ensure a localised failure -- like that we simulate here -- results in motions sufficiently widespread to implant a large-scale twist in the magnetosphere.

Electron MHD is, mathematically, a clean model for neutron-star crustal evolution. The range of input parameters is mostly quite constrained, results are insensitive to initial conditions, and the natural next step towards more realistic evolutions has typically been quite clear. Many of the considerable challenges have instead been numerical: for example, dealing with the formation of shocks due to Hall drift \citep{vigano12}, or moving from two to three spatial dimensions \citep{wood15,gour16}. Plastic flow introduces significant new uncertainties into modelling, and complicates the association of observables with the crust's underlying physics. We have limited understanding of how and when the crust fails and enters the plastic phase, and of the material response of the crust to high stress. The initial conditions and the crust's seismic history become important. On the other hand, this new physics is rich and interesting, and provides us with a possibility of using observations to constrain some of these unknown mechanical properties of the neutron-star crust.

%%%%%%%%%%%%%%%%%%%%%%%%%%%%%%%%%%%%%%%%%%%%%%%%%%%%%%%%%%%%
%%%%%%%%%%%%%%%%%%%%%%%%%%%%%%%%%%%%%%%%%%%%%%%%%%%%%%%%%%%%
%%%%%%%%%%%%%%%%%%%%%%%%%%%%%%%%%%%%%%%%%%%%%%%%%%%%%%%%%%%%

\section*{Acknowledgements}

We thank Chris Thompson \skl{and the anonymous referee for comments which helped to clarify the presentation of this work}.
SKL acknowledges support from the European Union's Horizon 2020
research and innovation programme under the Marie Sk\l{}odowska-Curie
grant agreement No. 665778, via fellowship UMO-2016/21/P/ST9/03689 
of the National Science Centre, Poland. Both authors thank the PHAROS COST Action (CA16214) for partial support.

This work used the DiRAC@Durham facility managed by the Institute for 
Computational Cosmology on behalf of the STFC DiRAC HPC Facility
(www.dirac.ac.uk). The equipment was funded by BEIS capital funding via 
STFC capital grants ST/P002293/1, ST/R002371/1 and ST/S002502/1,  Durham
University and STFC operations grant ST/R000832/1. DiRAC is part of the 
National e-Infrastructure.

%%%%%%%%%%%%%%%%%%%%%%%%%%%%%%%%%%%%%%%%%%%%%%%%%%%%%%%%%%%%%%%%%%
%%%%%%%%%%%%%%%%%%%%%%%%%%%%%%%%%%%%%%%%%%%%%%%%%%%%%%%%%%%%%%%%%%

\bibliographystyle{mnras}

%\newpage

\bibliography{references}

\label{lastpage}

\end{document}